\documentclass[preprint,eqsecnum,aps,floatfix,amsmath,amssymb,showpacs]{revtex4}
\usepackage{graphicx}
\usepackage{dcolumn}
\usepackage{bm}

\begin{document}
\newcommand\ie{{\rm i.e.\/}}
\newcommand\eg{{\rm e.g.\/}}
\newcommand\etal{{\it et al.\/}}
\newcommand\sign{\hphantom{$-$}}
\newcommand\Keff{K_{eff}}
\newcommand\Atotal{A_{\rm total}}
\newcommand\Ftotal{F_{\rm total}}
\newcommand\Fbulk{F_{\rm bulk}}
\newcommand\Meq{M_{\rm eq}}
\newcommand\Mref{M_{R}}
\newcommand\Minf{M_\infty}
\newcommand\hinf{h_\infty}
\newcommand\htilde{\tilde{h}}
\newcommand\rinf{r_\infty}
\def\t{\theta}
\newcommand\tinf{\theta_\infty}
\newcommand\tmax{\theta_{max}}
\newcommand\Sbr{\Sigma_{\beta|\gamma}}
\newcommand\Sab{\Sigma_{\alpha|\beta}}
\newcommand\Sabr{\Sigma_{\alpha|\beta\gamma}}
\newcommand\Sar{\Sigma_{\alpha|\gamma}}
\newcommand\F{{\cal F}}
\newcommand\DF{\Delta{\cal F}}
\newcommand\A{{\cal A}}
\newcommand\g{{\cal G}}
\def\D#1#2{\frac{d#1}{d#2}}
\def\PD#1#2{\frac{\partial#1}{\partial#2}}
\def\Y#1{Y^\infty_{#1}}
\def\Z#1{Z^\infty_{#1}}
\newcommand\mt{\widetilde m}
\newcommand\thc{\theta_c}
\newcommand\Mdot{{\buildrel\kern0.15em\scriptscriptstyle\bullet\over M}}

\newcommand\celc{\hbox{\thinspace$^\circ\hbox{\rm C}$}}
\newcommand\stu{\hbox{\hbox{erg}/\hbox{cm}$^2$}}
\newcommand\kelvin{\hbox{K}}
\newcommand\gcm{\hbox{g}/\hbox{cm}^3}
\newcommand\Ao{\thinspace{\buildrel \kern0.15em\scriptstyle\circ \over A}}

\newcommand\rmin{r_{\hbox{\ninerm min}}}

\newcommand\gl{\buildrel > \over {_<}}
\newcommand\vecr{\hbox{\bf r}}


\title{Scaling for Interfacial Tensions near Critical Endpoints}

\author{Shun-yong Zinn}%

\author{Michael E. Fisher}%

\affiliation{%
Institute for Physical Science and Technology\\
University of Maryland, College Park, MD 20742}%
\date{9 August 2004}

\begin{abstract}
Parametric scaling representations are obtained and studied for the asymptotic
behavior of interfacial tensions in the \textit{full} neighborhood of a fluid
(or Ising-type) critical endpoint, \ie, as a function \textit{both} of
temperature \textit{and} of density/order parameter \textit{or} chemical
potential/ordering field.  Accurate \textit{nonclassical critical exponents}
and reliable estimates for the \textit{universal amplitude ratios} are
included naturally on the basis of the ``extended de Gennes-Fisher''
local-functional theory.  Serious defects
in previous scaling treatments are rectified and complete wetting behavior
is represented; however, quantitatively small, but unphysical residual
nonanalyticities on the wetting side of the critical isotherm are smoothed
out ``manually.''  Comparisons with the limited available observations are
presented elsewhere but the theory invites new, searching experiments and
simulations, \eg, for the vapor-liquid interfacial tension on the two sides of
the critical endpoint isotherm for which an amplitude ratio $-3.25 \pm 0.05$ is
predicted.
\end{abstract}

\pacs{68.05.-n, 64.60.Fr, 64.70.Fx, 05.70.Jk\\
\vskip 180pt
To appear in Physical Review E}

\maketitle

\section{Introduction and Scaling Theory}
Consider, for concreteness, a binary liquid mixture consisting of two species,
$A$ and $B$.  For a full thermodynamic description, one needs three field
variables, say~$(T,h,g)$~\cite{Rowlinson, Fisher91b}.  The ordering field~$h$ is
conjugate to the order parameter, $M$.  For fluids, the order parameter
may (in leading order) be taken as the number density
$\Delta\rho = \rho - \rho_0(T,g)$ measured relative to a coexistence value
$\rho_0(T,g)$.  Alternatively, $M$ could be a composition variable such
as mole fraction difference, a volume fraction difference, and so forth.
For the nonordering field~$g$, one may take the pressure, or the chemical
potential of one species, either~$A$ or~$B$, etc.

\begin{figure*}
\includegraphics[scale=0.8]{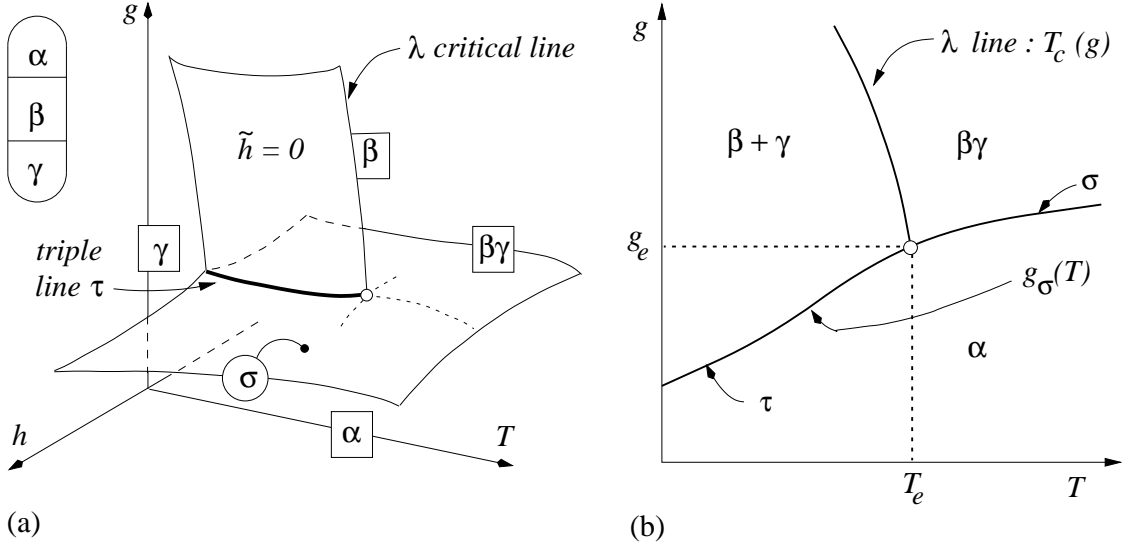}
\caption{\label{fig4:phase}
(a) Phase diagram of a binary liquid
mixture in the three-dimensional field space $(T,h,g)$: see text for
details.  (b) Section of the phase diagram (a) containing the plane $\htilde=0$.  The
critical endpoint, $(T_e,g_e)$, is where the critical line~$\lambda$
terminates at the first order transition line $g_\sigma(T)$.}
\end{figure*}

Fig.~\ref{fig4:phase}(a) illustrates a typical phase diagram in the
three-dimensional field space~\cite{Fisher91b}.  At any point on the surface
labeled~$\htilde = 0$, the system exhibits phase
separation into two coexisting phases, $\beta$ and~$\gamma$, rich in~$A$
and~$B$, respectively.  We will adopt the convention that the~$\gamma$ phase
has the higher (mass) density and hence sits at the bottom of a container when a
gravitational field is present: see the inset in Fig.~\ref{fig4:phase}(a).
By increasing temperature while keeping~$\htilde \equiv h - h_0(T,g)=0$,
the state point will reach the line~$\lambda$ which is a locus of critical
points,~$T_c(g)$.  Further temperature increase results
in mixing of the~$\beta$ and~$\gamma$ phases into a single phase,
say~$\beta\gamma$.  On the other hand, decreasing~$g$ at fixed $T < T_c(g)$ on
the $\htilde = 0$ surface leads to a triple-point line, $\tau$, at which
appears a new, noncritical or `spectator' phase $\alpha$ which represents the common vapor
of the liquid phases $\beta$, $\gamma$, and $\beta\gamma$: see the inset in Fig.~\ref{fig4:phase}(a).  A first-order transition,
between the vapor and the liquid phases, occurs across the vapor-pressure
surface labeled~$\sigma$ which meets the $\htilde = 0$ surface at the triple line.
The critical line, $\lambda$, and the triple point line, $\tau$, terminate
at a point $(T_e, 0, g_e)$: that is the ``critical endpoint.''


Recent field-theoretic renormalization group theory has confirmed explicitly
that the critical behavior at a critical endpoint is the same as
on the critical locus~\cite{Diehl2000, Diehl2001}.  Nevertheless, further,
new bulk thermodynamic singularities do appear at a critical
endpoint~\cite{Fisher90a, Fisher90b, Wilding97}.

Beyond the bulk, however,
there are singularities in \textit{interfacial} or \textit{surface}
\textit{tensions} when, in the presence of the vapor~$\alpha$, the two
phases~$\beta$ and~$\gamma$ merge into the homogeneous phase~$\beta\gamma$,
or vice versa~\cite{Fisher90a, Fisher90b}.  In Fig.~\ref{fig4:phase}(a)
one may follow the triple line and its smooth extension on the $\sigma$
surface, beyond the critical endpoint, or, in Fig.~\ref{fig4:phase}(b), simply
trace the line $g_\sigma(T)$.  The ``critical surface tension'' between
the coexisting phases vanishes below the critical endpoint temperature
$T_c = T_e$, as
\begin{equation}
\Sbr(T) \approx K |t|^\mu, \quad t\equiv(T-T_e)/T_e\to0-, \quad(\htilde=0),
\label{eq4.1:dsbr}
\end{equation}
where, via standard scaling relations~\cite{Rowlinson, Fisher90a, Zinn98,
Zinn99, Widom}, the critical exponent is given by $\mu = 2-\alpha-\nu$
so that $\mu \simeq 1.26$ for typical, three-dimensional fluids.
The amplitude $K$ has dimensions of energy per unit area where, here and below,
we adhere to the notation set out in the Appendix of Ref.~8. 
The ``noncritical tensions,'' $\Sabr$ and $\Sab$, should behave
~\cite{Rowlinson, Fisher90a}, after subtraction of a suitable, nonsingular
common background, $\Sigma_0(T)$, as
\begin{eqnarray}
\Delta\Sabr &\approx& K^+ |t|^\mu, \quad t\to0+, \quad(\htilde=0),
\label{eq4.1:dsabr}\\
\Delta\Sab &\approx& K^- |t|^\mu, \quad
t\to0-, \quad(\htilde = 0-), \label{eq4.1:dsab}
\end{eqnarray}
which relations serve to define the amplitudes $K^+$ and $K^-$.
Beneath $T_c$ the $\gamma$ phase coexists with the $\alpha$ and $\beta$ phases
and, in a sealed container, it sits below the $\beta$ phase
owing to its presumed 
heavier density.  Now consider, hypothetically, bringing into contact
the two phases
$\alpha$ and $\gamma$, near $\beta\gamma$ criticality; this produces
a new interface.
The corresponding noncritical
surface tension $\Sar$ can be obtained from Antonow's rule~\cite{Rowlinson}
which states
\begin{equation}
\Sar(T) = \Sab(T) + \Sbr(T). \label{eq4.1:Antonow}
\end{equation}
This relation can
be derived by supposing that the three phases $\alpha$, $\beta$, and $\gamma$
can coexist and meet with nonzero contact angles, and then by
letting the contact angle between
the interfaces $\alpha|\beta$ and $\beta|\gamma$ go to zero~\cite{Rowlinson}.
(It should be recalled, however, that Antonow's rule typically fails at
lower temperatures, specifically below a wetting temperature~$T_W$.)
 
Using the classical van der Waals or Cahn-Hilliard theory~\cite{Cahn} and a
model free energy of the Landau-expansion type, Widom~\cite{Widom77}
has studied various
properties of the noncritical interfaces, such as that between~$\alpha$
and~$\beta$, near the critical endpoint.  Later,
nonclassical critical exponents were embodied into the local free energy
expression via
postulated scaling forms~\cite{Widom}.  However, the original theory of
Widom and Ramos-G\'omez~\cite{Widom} led to an unexpected type of
correction in the surface tension, namely a~$|t|^\gamma$ term:
this is \textit{more} \textit{singular}
than the nonanalytic leading term~$|t|^\mu$ whenever the spatial
dimensionality, $d$, exceeds $3-\eta$~\cite{Fisher90a, Fisher90b}
which cannot be considered acceptable.

Fisher and Upton~\cite{Fisher90a} pointed out that, near the critical
endpoint, the amplitude ratios
\begin{equation}
P \equiv (K^+ + K^-)/K, \quad Q \equiv K^+/K^-, \label{eq4.1:PQ}
\end{equation}
should be \textit{universal}.  They reported 
mean-field calculations~\cite{Widom, Fisher90a, Fisher90b} yielding
\begin{equation}
P = -\hbox{$\frac{1}{2}$}(\sqrt{2} - 1) = -0.207\thinspace10\ldots\thinspace, \quad
Q = -\sqrt{2}, \label{eq4.1:PQmf}
\end{equation}
which should be valid for $d>4$.  However, to obtain more realistic values
for $d=3$, Fisher and
Upton~\cite{Fisher90a, Fisher90b} presented preliminary
calculations using an extended de~Gennes-Fisher (EdGF) local functional theory
\cite{Fisher90b} for fluid interfaces combined with
a simple ``interpolated linear model''
for the equation of state (as described in Ref.~9). 
This approach provided the significantly different estimates
\begin{equation}
P \simeq 0.1_2, \quad Q \simeq -0.83. \label{eq4.1:PQfu}
\end{equation}
More recently, the EdGF theory has also been applied to critical adsorption
problems~\cite{Upton2001}.

Our aim here, apart from estimating these universal ratios more precisely, is to
calculate the noncritical surface tension in \textit{nonzero}
\textit{ordering} \textit{field},
\ie, $\Delta\Sabr$, $\Delta\Sab$, and $\Delta\Sar$ as a function of $t$
\textit{and} $h$ (or the order parameter, $M$), on the~$\sigma$ surface
in the \textit{full} vicinity of a critical endpoint.
This is a basic problem of interfacial thermodynamics first broached
experimentally in pioneering work by Nagarajan, Webb and Widom~\cite{NWW}.
In fact, there is just a single function of two variables, $\Delta\Sigma(t,
h)$, that is to be sought once a suitable background $\Sigma_0(t,h)$ is
subtracted from the total surface tension, say $\Sigma_{\alpha|\bullet}(t,h)$.
Furthermore, in the first instance our main concern must be with the
singular, critical behavior which we may confidently expect to be described
in scaling form so that $\Delta\Sigma/|t|^\mu$ is related universally (for
$d<4$) to the scaled combination $M/|t|^\beta$ or, equivalently, to
$\tilde{h}/|t|^\Delta$, where, in standard notation, $\Delta = \beta +
\gamma = \beta\delta$.

Three decades after the development of renormalization group theory one
might expect this problem to be susceptible to such an approach.
Unfortunately, however, the still remaining difficulties, both technical and
conceptual, are profound despite the progress reported, for example, in the
review articles by D.~B.~Abraham, H.~W.~Diehl, and D.~Jasnow in Vol.~10 of
the series {\it Phase Transitions and Critical Phenomena} edited by C.~Domb
and J.~L.~Lebowitz (Academic, 1986) and subsequent developments (some of
which are referenced in further detail below).  Accordingly we report here
on calculations based on \textit{local} \textit{functional}
\textit{theories} going back historically to van der Waals analysis of the
critical surface tension $\Sigma_{\beta|\gamma}(T)$.  Specifically, we pick
up and develop the proposals of Fisher and Upton~\cite{Fisher90b} who
advanced, in particular, the EdGF theory which can consistently embody the
correct nonclassical critical point exponents, especially $\eta > 0$.

Such theories rely on the availability of an accurate description of the
\textit{bulk} thermodynamic properties.  To that end we will make heavy use
of the \textit{parametric} \textit{formulation} of scaling theory in the
neighborhood of a bulk critical point as extended to represent the true
correlation length, $\xi_\infty(T,h)$, and so provide a basis for
calculating interfacial tensions via local functional
theories~\cite{Fisher90b, Zinn98, Zinn99}.  For completeness and ease of
reference we recall the basic parametric expressions here.  As standard, one
first has
\begin{equation}
t = r k(\theta), \phantom{mmm} h = r^\Delta l(\theta), 
\phantom{mmm} M = r^\beta m(\theta),
\label{eq4.1:parametric}
\end{equation}
where $k(\theta)$ is an even function of the ``angular'' variable $\theta$
with $k_0 \equiv k(0) = 1$ and $k(\pm\theta_c) = 0$ so that $\theta = \pm
\theta_c$ corresponds to the critical isotherm, $T = T_c$, while $l(\theta)$
and $m(\theta)$ are odd, with $l(0) = m(0) = 0$ and $l(\pm\theta_1) = 0$
with $m(\theta_1)>0$ so that $\theta=\pm\theta_1$ describes the coexistence
surface $\tilde{h}=0$ beneath $T_c$: see Fig.~2 of~\cite{Zinn99}.

For general thermodynamic purposes, however, it proves more effective to
avoid integrating the equation of state to obtain the free energy;
accordingly~\cite{Zinn99}, we opt to treat $h$ [and $l(\theta)$] as
derived from the singular part of the reduced Helmholtz free energy which
may be written in scaling form as
\begin{equation}
{\cal A}_s(t, M) = r^{2-\alpha} n(\theta),
\label{eq4.1:As}
\end{equation}
where, following~\cite{Zinn99}, we can write
\begin{equation}
n(\theta) = r^{-2+\alpha} \thinspace \hbox{\rm sing}
	\left\{ \int_{M_R}^M h(M'; T) dM' \right\},
\label{eq4.1:ntheta}
\end{equation}
in which the operation \thinspace\thinspace sing\{$\bullet$\} \thinspace\thinspace extracts only the leading singular
part while $M_R < 0$ is a fixed reference value: see also (\ref{eq4.2:Adef}) below.  One
then finds that $l(\theta)$ is readily expressed in terms of $k(\theta)$ and
$n(\theta)$: see Eq.~(4.4) of~\cite{Zinn99}.

The generalized local functional theories proposed in \cite{Fisher90b} also
require the true correlation length, $\xi_\infty(T,h)$, which specifies the
exponential decay of correlations (in the presumed \textit{absence} of
long-range power-law or van-der-Waals-type interactions).  The corresponding
scaling form can be written~\cite{Zinn99}
\begin{equation}
\xi^2_\infty/2\chi = r^{-\eta\nu} a_\infty(\theta),
\label{eq4.1:xisqr}
\end{equation}
where $\chi = (\partial M/\partial h)_T$ is the reduced compressibility (or
susceptibility) while $\eta$ and $\nu$ are the standard correlation critical
exponents.  It is worth stressing that an essential feature of the
generalized local functional theories~\cite{Fisher90b} is to provide in a
consistent way for $\eta > 0$ since this is vital for the accurate
description of real and realistic model systems when $d < 4$.

The local functional theories, if they are to yield computations for the
\textit{interfacial} \textit{tension}, also require that ${\cal A}_s(M,T)$
and $\xi^2_\infty/2\chi$ are well defined \textit{through} the two-phase
region below $T_c$ where $|M|$ is less than $M_0(T)\approx B |t|^\beta$, the
order parameter at coexistence (or, in magnetic terms, the spontaneous
magnetization).  While this is most certainly questionable from a rigorous
viewpoint, one may in practice construct \textit{trigonometric}
\textit{forms} for $k(\theta)$, $m(\theta)$, $n(\theta)$, etc., which
extrapolate smoothly (and, indeed, analytically) to $|\theta| > \theta_1$ and
so through the two-phase region: see \cite{Fisher90b, Zinn98, Zinn99}.  In
such cases we take $k(\theta)$, $m(\theta)$, etc., as smooth periodic
functions, of appropriate parity, in the interval $-\theta_0 \leq \theta
\leq \theta_0$ where $\theta_0$ then corresponds to $h = M = 0$ for $T <
T_c$.  On this framework an `extended sine model' has been built and fitted
to reliable estimates of critical exponents and amplitude ratios for the
($d=3$)-dimensional Ising model~\cite{Zinn99}.  The resulting scaling
functions will be used here to study the interfacial tensions near a
critical endpoint.

In the scaling region the singular part of the full interfacial tension can
consequently be written parametrically as
\begin{equation}
\Delta\Sigma(t,h) = r^\mu s(\t), \label{eq4.1:sigma}
\end{equation}
and our basic task is to calculate the angular surface tension
function $s(\t)$.  Note that $\Delta\Sigma$ represents (i) $\Delta\Sabr$ when
$|\t| < \t_c$, (ii) $\Delta\Sab$ when $\t_c <
\t \le \t_1$, and (iii) $\Delta\Sar$ when
$-\t_1 \le \t \le -\t_c$, in accordance with Fig.~\ref{fig4:phase} and the notation explained
above.
Once $s(\t)$ is determined, the surface tension can also be written
in the standard scaling form
\begin{equation}
\Delta\Sigma \approx K |t|^\mu S_\pm ({\bar D} h/|t|^\Delta),
\label{eq4.1:Spm}
\end{equation}
where the universal scaling function $S_\pm(x)$
can be readily calculated.  (Note that, as customary, the
subscripts~$+$ and~$-$ stand for~$t\geq0$
and $t\leq0$, respectively.)  The amplitudes $K$ and $\bar D$ here are the
nonuniversal
metric factors needed for normalization and to make $S_\pm(x)$ and the
argument~$x$, dimensionless.  In the notation of \cite{Zinn98} we take
${\bar D} = C^+/B$ and $x = \tilde h$, which provided a
convenient normalized variable in the analysis of Ref.~9. 

Experimentally, as mentioned, Nagarajan, Webb, and Widom (NWW) \cite{NWW}
were the first to test theoretical predictions for
the universal surface-tension scaling functions off the $\tilde h = 0$ axis
in their studies of mixtures of isobutyric acid
and water.  For the same mixture, Howland, Wang and Knobler~\cite{Howland80}
measured the
critical surface tension $\Sigma_{\beta|\gamma}$ and Greer~\cite{Greer76} measured
densities on the coexistence curve.  The two latter experiments can be used
to provide a 
consistency check and calibration of the NWW data~\cite{Zinn97}.
Other mixtures have also been examined.
Quasi-binary mixtures of {\it n}-octadecane and {\it n}-nonadecane in ethane
have been studied by Pegg et al.~\cite{Pegg85}
to measure the surface tensions through and near both the
upper and the lower critical endpoints.  The
surface tension of the water and 2,5-lutidine system at and off the critical
composition has been measured by Amara et al.~\cite{Amara91}.  For a similar
mixture of water and 2,6-lutidine, Mainzer-Althof and Woermann determined the
values of $P$ and $Q$ experimentally~\cite{Mainzer}.  Interfacial tensions of
the critical mixture of 2-butoxyethanol and water have been measured
by Ataiyan and Woermann~\cite{Woerman95}.  The applications of the present
theory to these various data will be presented elsewhere~\cite{FisherNote}.

The rest of this article proceeds as follows.  In Sec.~II, we
review briefly the classical theory of interfaces.  The construction of more
general local free energy functionals is taken up in Sec.~III.
Following
Fisher and Upton~\cite{Fisher90a, Fisher90b} we work out the details of the
extended de~Gennes-Fisher (EdGF) ansatz and obtain formulae for the equilibrium
order parameter profile and the surface tension.  The hypothesis that the
noncritical vapor phase~$\alpha$ can be replaced by a wall with a surface
field~$h_1$~\cite{Fisher90a, Fisher90b} is introduced in Sec.~IV.
In the scaling limit $h_1/|t|^{\Delta_1} \to -\infty$ (where $\Delta_1$ is
the appropriate \textit{surface} critical exponent), there appear terms in
the total wall tension
that diverge although remaining analytic in~$t$.
After subtracting these divergent terms,
we can express the finite singular part of the surface tension
near a critical endpoint explicitly in parametric scaling form.
In Sec.~V these expressions are evaluated numerically for $d = 3$.
However, an unphysical although quite small \textit{cusp} is uncovered
in the basic scaling function $s(\theta)$ in (\ref{eq4.1:sigma}).  Its origin
is discussed and found to reside in a fairly subtle deficiency of the EdGF
scheme.  By using an interpolation scheme, the
cusp can be smoothed out leading to acceptable approximations
for the universal scaling functions $S_\pm(x)$ in (\ref{eq4.1:Spm}).  On this basis
various concrete numerical results are presented in Sec.~VI for the surface
tensions in the vicinity of a critical endpoint.  Sec.~VI contains some
brief concluding remarks.

\section{Local Functional Theories For Fluid Interfaces}
\subsection{The Auxiliary Free Energy Function}
Let us consider various free energies that will be needed in
discussing the local functional theory of fluid interfaces in a general
way.  Let $A(M,T)$ be the true equilibrium
Helmholtz free energy density that preserves the
appropriate convexity properties in~$M$ and~$T$ \cite{Callen}.
The free energy density $A(M,T) = \Atotal(M,T)/V$
can be obtained by integrating the equation of state,
\begin{equation}
A(M,T) = \int^M_{\Mref} h(M')\thinspace dM' + A(\Mref,T),
\label{eq4.2:Adef}
\end{equation}
using a fixed reference value $\Mref \ne 0$: such a choice of
reference value guarantees no singularity across $t=0$ in
$A(\Mref,T)$.  In order to keep track of dimensions, we take $M$ as the
density difference $(\rho-\rho_c)$ from now on.

Now let $A^\dagger(M,T)$ be the
Helmholtz free energy density in the one-phase region and its presumed analytic
continuation into the multiphase region~\cite{Fisher90b, Zinn98, Zinn99}.  The typical van der Waals loop,
which does not respect the convexity, should appear in the multiphase region of
$A^\dagger(M,T)$.  The Maxwell construction applied to $A^\dagger(M,T)$ repairs
the convexity (although this is, of course, {\it ad hoc\/}).  Evidently,
$A(M,T) = A^\dagger(M,T)$ outside the multiphase region 
while $A^\dagger(M,T) \ge A(M,T)$
inside.  The excess free energy $A^\dagger(M,T) - A(M,T)$ which is then
always nonnegative, serves in the local functional theories, to determine
the structures of the interfaces between the coexisting phases~\cite{Widom72}.

The conjugate free energy density $F(\hinf,T) = \Ftotal(\hinf, T)
/V$, where the subscript
$\infty$ denotes a bulk equilibrium quantity, may be obtained from
$A(M,T)$ via the Legendre transform,
\begin{eqnarray}
F(\hinf;T) &\equiv& \min_M[A(M,T) - \hinf M], \label{eq4.2:Fdef}\\
	&=& A(\Minf,T) - \hinf \Minf,
\end{eqnarray}
where $\Minf = M(T,\hinf)$ is the bulk equilibrium value of the order parameter.

The interfacial tension that we aim to calculate is the
excess free energy of a system in equilibrium created by one or more
interfaces.
All local functional theories for interfaces introduce an auxiliary free
energy $W[M(z);T,\hinf]$ that resembles the excess free energy: $W(M)$ is
always nonnegative and vanishes only when the profile, $M(z)$, takes an equilibrium
value of the order parameter corresponding to one of the coexisting phases
($M_\alpha$ for the $\alpha$ phase, $M_\beta$ for $\beta$, etc.).
Since the equilibrium values of the order parameter $M_\alpha$, $M_\beta$,
etc., are specified both by $T$ and $\hinf$, the dependence of the auxiliary free
energy on $T$ and $\hinf$ must not be overlooked.

Thus the auxiliary free energy function $W(M;T,\hinf)$
needed in a local functional theory can be defined
by~\cite{Fisher90a, Fisher90b}
\begin{eqnarray}
W(M;T,\hinf) &\equiv& A^\dagger(M,T)-\hinf M-F(\hinf, T), \label{eq4.2:Wdef}\\
	&=&A^\dagger(M,T) - A^\dagger(\Minf,T) - \hinf (M-\Minf),
		\label{eq4.2:Wdef2}
\end{eqnarray}
where we have used $A^\dagger(\Minf) = A(\Minf)$.  Since $A^\dagger(M)$ represents
the analytic continuation of $A(M)$, we may also use
(\ref{eq4.2:Adef}) to write
\begin{equation}
W(M;T,\hinf) = \int^M_{\Minf} h(M',T) \thinspace dM' - \hinf (M - \Minf),
\label{eq4.2:Wrep}
\end{equation}
because the equation of state must then equally have an analytic
continuation: notice that $A(\Mref,T)$ cancels out.  From
this one 
immediately finds that $W(\Minf;T,\hinf) = 0$ and $(\partial W/\partial
M)_{M=\Minf} = 0$.  Starting from $A^\dagger(M,T) \ge A(M,T)$ and using
(\ref{eq4.2:Fdef}) and (\ref{eq4.2:Wdef}) one obtains
\begin{equation}
W(M;T,\hinf) \ge A(M,T) - \hinf M - \min_M[A(M,T) - \hinf M] \ge 0.
\end{equation}
Thus, $W(M;T,\hinf)$
vanishes at the equilibrium values of $M$ for all $T$ and $\hinf$ and its
leading term in the expansion about $\Minf$ is quadratic in $(M-\Minf)$.
Otherwise, $W$ takes only positive values.

\subsection{van der Waals Theory}
It is helpful to review briefly the van der Waals theory of interfaces which is a Landau-type
classical theory~\cite{Rowlinson, Cahn, Davis}.  Assuming the existence of the
local free energy $W[M(z);T,\hinf]$ that can be expanded in powers of $M$ and
$T-T_c$, the van der Waals theory takes the local excess
free-energy density functional as a sum of two terms~\cite{Rowlinson},
namely,
\begin{equation}
{\mit\Delta} f[M(z)] = W[M(z)] + \hbox{$\frac{1}{2}$} J_0 \left(\frac{dM}{dz}\right)^2,
\label{eq4.2:mfDf}
\end{equation}
where $J_0 \thinspace(= \xi^2/\chi)$ is a constant and $z$ is the perpendicular distance from the
interface presumed to be flat.  The square-gradient term, $(dM/dz)^2$,
accounts for spatial inhomogeniety in the simplest manner.
The overall excess free energy is then
given by a volume integral of~${\mit\Delta} f(z)$.  Translational
invariance parallel to the interface enables one to factor
out the area in the volume integral so that the free energy per unit area, or
surface tension, can be written as \cite{Rowlinson}
\begin{equation}
\Sigma[M(z)] = \int dz\thinspace {\mit\Delta} f[M(z)].
\end{equation}
Functional minimization of $\Sigma[M]$ with respect to $M(z)$ yields a
differential equation for the equilibrium order parameter profile $M(z)$.
To supply boundary conditions let us consider, for example, a system
containing two bulk
equilibrium phases with $M = M_\beta$ (or $M_{-\infty}$) and $M_\gamma$
(or $M_{+\infty}$) located at $z=-\infty$ and
$+\infty$, respectively.  For convenience, $M_\beta$ and $M_\gamma$
can be taken equal in magnitude but opposite in sign ($M_\gamma > 0$).
Then, near criticality the equilibrium order parameter profile behaves
like $M_\gamma \tanh(z/\xi)$ for the stated boundary conditions and the
resulting surface tension is
\begin{equation}
\Sigma = \int^{M_\gamma}_{M_\beta} dM \thinspace \sqrt{2 J_0 W(M)}.
\end{equation}
With a suitable representation for $W[M]$, one can also study the critical
and noncritical surface tensions~\cite{Widom, Widom77}, as well as the
critical wetting transition~\cite{Cahn77}.  However, this classical
square-gradient theory \textit{cannot} satisfactorily embody all the correct critical exponents.  This is
because the square-gradient term in (\ref{eq4.2:mfDf}), when the analysis is adapted
to study the
decay of correlations (rather than just the overall interfacial free energy),
always implies the exponent value $\eta = 0$ that
is valid only for $d \ge 4$.  Thus even upon using an equation of state
that obeys scaling and embodies correct exponents (such as the extended
sine model discussed in~\cite{Zinn99}), a square-gradient
form for the local free-energy yields the classical value $\eta=0$ for
the correlation function decay~\cite{FisherEta}.

\section{Fisher-Upton Theory For Fluid Interfaces}
In order to generalize the classical square-gradient theory, we start with
the total free energy of the binary mixture in the following form
\begin{equation}
\Ftotal(T,\hinf,g) = \min_{M({\vecr})} {\F[M]},
\end{equation}
where $\F[M;T,\hinf]$ is a sought-for local free energy functional.  Assuming
translational invariance along directions parallel to the interfaces,
we may simply take $M({\vecr}) = M(z)$.  The solution $M(z)$ that minimizes
the functional $\F[M]$ is the equilibrium order parameter profile
for the specified $T$, $\hinf$ and boundary conditions.  We can
suppose that $\F[M]$ has a homogeneous part describing the uniform bulk
phases and an inhomogeneous part $\DF$, so that
\begin{equation}
\F[M(z)] = \Fbulk(T,\hinf,g) + \DF,
\end{equation}
where $\Fbulk(T,\hinf,g)$ is the bulk free energy (and $g$
is the third field: see Fig.~\ref{fig4:phase}).  If there
were no interfaces, we would have $\DF=0$, and $\Ftotal
= \Fbulk$.  Following~\cite{Fisher90b}, we
now consider a general local free energy functional of the form
\begin{equation}
\DF = \int d\vecr \thinspace \A(M,\Mdot; T, \hinf, g), \label{eq4.3:Adef}
\end{equation}
where $\Mdot \equiv dM/dz$.  Without great loss of generality, we may
write~\cite{Fisher90b}
\begin{equation}
\A(M,\Mdot;T,\hinf,g) = W(M)[1 + J(M)\g(\Lambda\Mdot)],
\label{eq4.3:Aeq}
\end{equation}
where $J$ and $\Lambda$ are to be functions of $M$ and $T$.
The function  $\g(x)$ should be
even since the sign of $\Mdot$ cannot matter in the free
energy.  Now $\A$ should vanish when $\Mdot = 0$ so that $\DF=0$; this
implies $\g(0)=0$.  To determine other properties of $\g(x)$ we must proceed further.

For semi-infinite cases where an external wall is located at $z=0$, the
postulate~(\ref{eq4.3:Adef}) must be modified by adding a boundary
term~\cite{Nakanishi}
\begin{equation}
\frac{\DF[M]}{A} = \int^\infty_0 dz\thinspace \A(M,\Mdot) + f_1(M_1;h_1),
\label{eq4.3:DF}
\end{equation}
where $A$ is the area of the interface.
The surface field $h_1$ acts only at $z=0$ and is coupled to $M_1 \equiv
M(z\negthinspace=0\negthinspace)$, the order parameter on the boundary.

Now let us minimize
\begin{equation}
\frac{\DF[M(z)]}{A} = 
\int^\infty_0 dz \thinspace [\A(M,\Mdot) + \delta(z)f_1(M)]
\end{equation}
with respect to $M(z)$ where the boundary term has been
absorbed into the integral.
Then in the usual way, integrating by parts, one obtain
\begin{eqnarray}
\frac{\delta\DF[M(z)]}{A}
    &=& \int^\infty_0 dz\thinspace \left[\left\{\PD{\A}{M}
	- \D{}{z}\left(\PD{\A}{\Mdot}\right)\right\}\delta M 
	+ \delta(z)\left\{\D{f_1}{M}-\PD{\A}{\Mdot}\right\}\delta M\right]\nonumber\\
    &\quad&\hskip 5 em +\PD{\A}{\Mdot}\delta M\Bigg|_{z=\infty},\label{eq4.3:step2}
\end{eqnarray}
and hence finds
\begin{eqnarray}
\PD{\A}{M} - \D{}{z}\left(\PD{\A}{\Mdot}\right) &=& 0,
	\label{eq4.3:sol1}\\
\left(\PD{\A}{\Mdot} - \D{f_1}{M}\right)_{z=0} &=& 0,
	\label{eq4.3:sol2}
\quad\hbox{\rm and}\quad\quad
\PD{\A}{\Mdot}\Bigg|_{z=\infty} = 0.
\end{eqnarray}
However, since $\A(M,\Mdot)$ has no explicit $z$-dependence, we can
integrate (\ref{eq4.3:sol1}) to get the first-order differential equation
\begin{equation}
\A(M,\Mdot) - \Mdot\PD{\A}{\Mdot} = {\cal C},\label{eq4.3:eom}
\end{equation}
in which $\cal C$ is a constant.

Now consider a fully infinite situation so that the lower limit in (\ref{eq4.3:DF}) becomes
$-\infty$ and the surface term drops out.
Functional minimization again yields (\ref{eq4.3:sol1}), and thence (\ref{eq4.3:eom}), while the
first (wall) condition in (\ref{eq4.3:sol2}) becomes simply the bulk condition
$(\partial\A/\partial\Mdot)_{z=-\infty} = 0$.

In the semi-infinite situation, the form~(\ref{eq4.3:Aeq}) leads to
\begin{equation}
W(M_1)J(M_1)\Lambda(M_1)\g'(\Lambda\Mdot)\Big|_{z=0} = \D{f_1}{M_1},
\label{eq4.3:M1}
\end{equation}
which is an equation determining $M_1$.  Similarly, the two bulk conditions
yield
\begin{equation}
\PD{\A}{\Mdot}\Bigg|_{z=\pm\infty} = W(M_{\pm\infty}) J(M_{\pm\infty})
 \Lambda(M_{\pm\infty})
    \g'(\Lambda\Mdot_{\pm\infty}) = 0;
\end{equation}
but since $W(M_{\pm\infty})=0$ these conditions should hold
automatically provided that $J(M_{\pm\infty})$,
$\Lambda(M_{\pm\infty})$, and $\g'(\Lambda\Mdot_{\pm\infty})$ do not diverge.
We will see below that these
functions have nice behavior so that we may
forget the bulk conditions.

Far away from the wall or the interfaces,
in the infinite and semi-infinite cases, one should see only the bulk behavior
of the system.  Hence, it is natural to require
\begin{equation}
M(z) \to M_{\pm\infty}, \quad\Mdot(z) \to 0,
\qquad\hbox{\rm as}\quad |z|\to\infty.
\end{equation}
Because $\A(M_{\pm\infty},0) = 0$ the constant $\cal C$ in the first
integral~(\ref{eq4.3:eom}) must thus vanish.  Then using the postulate~(\ref{eq4.3:Aeq})
one is led to
\begin{equation}
x \equiv \Lambda(M)\Mdot =
  \frac{1+J(M)\g(x)}{J(M)\g'(x)}, \label{eq4.3:prof}
\end{equation}
which represents a differential equation for the equilibrium profile $\Meq(z)$.

In order to devise appropriate expressions for $\A(M,\Mdot)$ which
generalize the earlier de~Gennes-Fisher theory \cite{Fisher78}, applicable
only at $T=T_c$, Fisher and Upton \cite{Fisher90b} introduced the
following desirable physical features or desiderata:

\noindent\hskip 3 em{\bf A}. The correct, non-classical critical exponents
should be embodied, both in $d=2$ and in $d=3$.
Conversely, a reduction to van der Waals or Landau theory
should be implied
whenever the classical critical exponents are assumed.

\noindent\hskip 3 em{\bf B}. Near the critical point, all thermodynamic
functions should satisfy the scaling and analyticity requirements: in
particular, singularities (or nonanalyticities) should appear only at
bulk criticality, i.e., when $t = h = \Minf = 0$.

\noindent\hskip 3 em{\bf C}. For the semi-infinite system, the
critical-point decay of the profile should behave in accordance with
scaling as \cite{Fisher78, Dietrich95}
\begin{equation}
M_c(z) \sim z^{-\beta/\nu}. \label{eq4.3:mcdecay}
\end{equation}

\noindent\hskip 3 em{\bf D}. When two plates separated by a finite
distance $L$ are immersed
in a critical system, the order parameter decay~(\ref{eq4.3:mcdecay}) should
have a correction factor so that
\begin{equation}
M_c(z) \sim z^{-\beta/\nu} \left[1 + j_2 \left(z/L\right)^{d^*} +
	\cdots \right], \quad L\to\infty,
\end{equation}
where $d^* = (2-\alpha)/\nu$ and $j_2$ is some coefficient.  This prediction
of the de~Gennes-Fisher theory \cite{Fisher78} has been
verified by several analyses including exact Ising model
calculations for $d=2$~\cite{Fisher90a, Fisher90b} and field-theoretic calculations
in $\epsilon = 4 - d$ dimensions~\cite{Dietrich93, Krech95, Dietrich96}.

\noindent\hskip 3 em{\bf E}. Away from criticality, the order parameter
should always decay exponentially
\begin{equation}
\Delta M(z)\equiv M(z) - \Minf \sim e^{-z/\xi}, \quad z\to\infty,
\label{eq4.3:bfE}
\end{equation}
where, clearly, $\xi(T,h)$ denotes the \textit{true} correlation length,
$\xi_\infty$, but for brevity we will neglect the subscript $\infty$.  (Note
again that we are assuming the absence of power-law forces or the likelihood
that, if present, they do not enter explicitly into the asymptotic scaling
functions.)

\noindent\hskip 3 em{\bf F}. For a finite critical slab, in the same
situation as {\bf D}, the critical profile will exhibit a \textit{minimum}
for similar boundary conditions satisfying $M(z\negthinspace=\negthinspace0)$,
$M(z\negthinspace=\negthinspace L) > 0$,
or a \textit{zero} for opposing boundary conditions $M(z\negthinspace=\negthinspace0)>0$
and $M(z\negthinspace=\negthinspace L)<0$
\cite{Fisher80}.  It is expected that this profile, $M_c(z)$, behaves
analytically near $z=z_0$ as
\begin{equation}
M_c(z) = k_1 (z-z_0)^{\lambda_1}[1 + k_2 (z-z_0)^{\lambda_2} + \cdots], 
\end{equation}
where {\bf F(i)} $\lambda_1=0$ for the similar case, or $\lambda_1 = 1$ for
the opposing case.  {\bf F(ii)} for the further exponents, one should have
$\lambda_2 = 2,\quad\lambda_4 = 4$, etc.

\noindent\hskip 3 em{\bf G}.  Away from criticality a square-gradient
expansion in the local free energy functional is expected to be correct
and so should be reproduced by a satisfactory theory.

\noindent\hskip 3 em{\bf H}.  To describe adsorption on a wall at $z=0$,
the theory should be consistent with the thermodynamic relation
\begin{equation}
\Gamma \equiv \int^\infty_0
\Delta M(z)\thinspace dz = -\left(\PD{\Sigma}{h}\right)_T, 
\end{equation}
where $\Sigma$ is the wall (or surface) free energy.

\noindent\hskip 3 em{\bf I}.  The order parameter profile $M(z;T,h,g;L)$
should be analytic in all noncritical regions.  (Compare with {\bf B}
above).

The starting point of the Fisher-Upton theory is the de~Gennes-Fisher (dGF)
ansatz~\cite{Fisher78} for $T=T_c$ which in the general expression (\ref{eq4.3:Aeq}) is
given by~\cite{Fisher90a, Fisher90b},
\begin{eqnarray}
&&J = \hbox{const.}, \quad\Lambda(M)=\xi(M)/M, \quad\g(x)=|x|^{2-\tilde\eta},
	\nonumber\\
&&\hbox{with}\quad \tilde\eta = 2\eta/(d^*+\eta).
\end{eqnarray}
This form satisfies {\bf A -- D} and {\bf F(i)}.  However, the dGF theory
applies only \textit{at} criticality.  The
extension proposed by Fisher and Upton (EdGF) uses $\chi(M,T) = (\partial M/\partial h)_T$ and postulates
\begin{equation}
J=1, \quad\Lambda(M;T,\hinf,g)=\sqrt{\xi^2(M,T)/2\chi(M,T)W(M;T,\hinf)}.
\label{eq4.3:edgf}
\end{equation}
In order to satisfy {\bf E}, one finds that the condition
\begin{equation}
\widehat\g(1)=1 \qquad \hbox{with} \qquad \widehat\g(x) \equiv x\g'(x) -
\g(x), \label{eq4.3:curlyg}
\end{equation}
must be satisfied.  Also, in order to satisfy {\bf F}, $\g(x)$ should behave as
\begin{eqnarray}
&&\g(x) = G_0 + G_\infty |x|^{2-\tilde\eta} [1+l_1 x^{-\tau} + l_2 x^{-2\tau} +
\cdots],\nonumber\\
&&\qquad\qquad\quad\hbox{with} \quad \tau=2\beta/(\beta+\nu).
\label{eq4.3:gexp8}
\end{eqnarray}
Finally, for small $x\to0$ the validity of a gradient expansion, {\bf G},
can be seen to require~\cite{Fisher90a, Fisher90b}
\begin{equation}
\g(x) = x^2 + G_2 x^4 + G_4 x^6 + \cdots.
\label{eq4.3:gexp0}
\end{equation}

Using the EdGF postulate~(\ref{eq4.3:edgf}) and the
condition~(\ref{eq4.3:curlyg}), the solution of the general profile
equation~(\ref{eq4.3:prof}) reduces to $x=\pm1$, or
\begin{equation}
\Mdot = \pm1/\Lambda(M), \label{eq4.3:edgfprof}
\end{equation}
where the signs $\pm$ must be chosen appropriately.  Hereafter, we take the
$+$ sign, for an increasing profile as $z\to+\infty$.
The wall free energy then follows from~(\ref{eq4.3:DF}) as
\begin{equation}
\Sigma = \DF[\Meq(z)]/A =
[1+\g(1)]\int^\infty_0dz\thinspace W(M) + f_1(M_1),\label{eq4.3:edgfsig2}
\end{equation}
where (\ref{eq4.3:Adef}), (\ref{eq4.3:edgf}), (\ref{eq4.3:edgfprof}) and
$\g(-x) = \g(x)$ have been used.  
Using~(\ref{eq4.3:edgfprof}) once again, one
can rewrite the semi-infinite integral as
\begin{equation}
\Sigma = [1+\g(1)]\int_{M_1}^{M_\infty} \hskip -1 ex dM \thinspace\thinspace
W(M) \Lambda(M) + f_1(M_1).\label{eq4.3:edgfsig}
\end{equation}

Now
the thermodynamic consistency condition, {\bf H},
leads to~\cite{Fisher90b}
\begin{equation}
1+\g(1)=2 \qquad \hbox{or} \qquad \g(1)=1.
\label{eq4.3:thermcond}
\end{equation}
It is remarkable that both the profile and the wall or
interfacial free energy, $\Sigma(T,h)$, do not depend on the details of
$\g(x)$~\cite{Fisher90b}.  However, one must note that the EdGF ansatz fails
to satisfy {\bf I} in certain situations~\cite{Fisher90b}.

In order to repair this last problem, a generalized de~Gennes-Fisher ansatz
(GdGF) was devised \cite{Fisher90b}.  It satisfies all of the desiderata
{\bf A-I}.  However, in contrast to the EdGF theory,
the profile equation, the surface tension
formulae, etc., now depend on $\g(x)$ explicitly.   In order to obtain
quantitative results, one must devise a representation for $\g(x)$ that
reproduces~(\ref{eq4.3:gexp8}) and an analogue of~(\ref{eq4.3:gexp0}).  One
can indeed achieve this; but the resulting calculations become considerably
more complicated than those for the EdGF theory.
Accordingly we have explored numerically only the EdGF formulation.

It must be recognized, however, that both EdGF and GdGF theories ignore
capillary-wave fluctuations of a free interface, which are important
for $d \le 3$ \cite{Fisher90a, Fisher90b}.  Also, as indicated above,
the analytic continuations of $W(M)$,
$\xi^2(M)/2\chi(M)$, etc., into the multi-phase region have no known meaning
in the strict sense of rigorous statistical mechanics.
However, we expect that both theories will produce reasonably reliable
results when fitted to exact $d=2$ and $d=4$ results and good estimates
for various $d=3$ parameters, since
they embody many correct physical features.

Indeed, Upton's $\thinspace\thinspace\epsilon = 4 - d\thinspace\thinspace$
  expansion results for the universal
amplitude ratio $\thinspace\thinspace Q = K^+/K^-$ \cite{Upton92} demonstrate this point quite
well.  Using the field-theoretic approach to surface critical phenomena,
he obtained the exact $\epsilon$-expansion $\thinspace\thinspace Q = -\sqrt{2} +
1.521\thinspace257\thinspace\epsilon + {\cal O}(\epsilon^2)$.
Then, using the EdGF theory and the linear parametric model
(which is known to be exact to order $\epsilon^2$ \cite{Brezin71, Brezin73}),
he found $\thinspace\thinspace Q = -\sqrt{2} + 1.522\thinspace96_2\epsilon + {\cal O}(\epsilon^2)$.
The coefficients of $\epsilon$ differ by only 0.1\%.

\section{EdGF Expressions for Surface Tension Near a Critical Endpoint}
We now discuss in more explicit detail the application of the EdGF
theory sketched above to the vicinity of a
critical endpoint.  Allowing for a boundary term, which will be discussed
further below, (\ref{eq4.3:edgfsig}) can be written as
\begin{equation}
\Sigma(T, h, g) = 2\int_{M_1}^{\Minf} dM \thinspace \sqrt{W(M)
  \xi^2(M)/2\chi(M)} + f_1(M_1), \label{eq4.4:Sigma}
\end{equation}
where the arguments $\hinf$ and $T$ are understood and~(\ref{eq4.3:edgf})
and (\ref{eq4.3:thermcond}) have been used.

\subsection{Scaling Forms}
To embody the appropriate nonclassical critical exponents and satisfy the
desiderata {\bf A} and {\bf B} we should, clearly, adopt scaling forms for
$W(M;T,\hinf)$ and the combination $\xi^2(M,T)/\chi(M,T)$.  As discussed, the
required expressions must continue analytically (or, at least, sufficiently
smoothly) into the two-phase region $|M| < M_0(T)$.  We should also recall
the necessity for including the further field $g$ and the lambda line
$T=T_c(g)$: see Fig.~\ref{fig4:phase}.  Following~\cite{Zinn99}, we thus introduce
the dimensionless asymptotic scaling variables
\begin{equation}
\widetilde m \equiv M/B|\tilde t|^\beta, \quad
\tilde h \equiv [h - h_0(T,g)]/(B/C^+)|\tilde t|^{\beta+\gamma},
\label{eq4.4:norm}
\end{equation}
where the reduced temperature deviation is now \cite{Fisher90a, Fisher90b}
\begin{equation}
\tilde t = [T - T_c(g)] / T_e,
\label{eq4.4:ttilde}
\end{equation}
while, for convenience, the tilde on $h$ now denotes the fully scaled field
and, for brevity, the phase boundary term, $h_0(T,g)$, will usually be
neglected below.  As usual~\cite{Zinn98}
$B = B(g)$ and $C^+ = C^+(g)$ are, respectively, the critical
amplitudes of the spontaneous order, $M_0(T,g) \approx B|t|^\beta$,
and the susceptibility, $\chi_0^+(T,g) = (\partial M/\partial h)_{T>T_c}
\approx C^+/t^\gamma$, on the $\htilde=0$ surface above $T_c(g)$.

Then, following \cite{Fisher90a, Fisher90b}, we can write the scaling form
\begin{equation}
\xi^2(M,T,g)/2\chi(M,T,g) \approx |M|^{-\eta\nu/\beta} Z_\pm(\mt), 
\end{equation}
which is crucial in going beyond van der Waals or Landau square-gradient
theory because it introduces the small but positive (for $d<4$)
exponent~$\eta$.  As above, the subscripts $+$ and $-$ will always
denote $\tilde t > 0$ or $<0$, respectively.
However, it must be realized here and below that the scaling functions
$Z_+$ and $Z_-$ and, likewise, others are, in fact, representations
of a \textit{single}, generally analytic scaling function continuing
smoothly through $t \gl 0$.  Thus, more explicitly, to ensure the analyticity
of $\xi^2/2\chi$ across the surface $T=T_c(g)$ or $\tilde t=0$ (recall {\bf
B}) the scaling functions $Z_\pm$ must have large $\widetilde m$ expansions
of the form
\begin{equation}
Z_\pm(\mt) = \Z0\left[1
  + \sum^\infty_{n=1} \Z{n} \left(\pm |\mt|^{-1/\beta}\right)^n\right].
\label{eq4.4:Zm}
\end{equation}
When $T \to T_c(g)\pm$ the terms in the sum clearly generate only the
integral powers ${\tilde t}^n$ as required by analyticity when $M \ne 0$.

The correct analyticity is most conveniently
incorporated by using the parametric representations of the scaling functions
recalled in Sec.~I.
The only new features required near a critical endpoint
are the replacement of $t$ by $\tilde
t$, as defined in~(\ref{eq4.4:ttilde}) and allowance for the (smooth)
dependence of the nonuniversal factors $m_0 \equiv m(0)$ and $l_0 \equiv
l(0)$ on the field~$g$: see (\ref{eq4.1:parametric}) and~\cite{dimen}.
With this understanding the parametric forms (\ref{eq4.1:parametric})--(\ref{eq4.1:ntheta}) 
will be adopted.  Then one can express
the coefficients entering~(\ref{eq4.4:Zm}) as
\begin{eqnarray}
\Z0 &=& a_\infty(\thc) [m(\thc)]^{\eta\nu/\beta}
	\equiv a_{\infty c} (m_c)^{\eta\nu/\beta},\\
\Z1 &=& \frac{[m_c/B]^{1/\beta}}{k'_c}
    \left\{{a'_{\infty c} \over a_{\infty c}} + \frac{\eta\nu}{\beta}
        {m'_c \over m_c} \right\},\\
\Z2 &=& \frac{[m_c/B]^{2/\beta}}{2[k'_c]^2}
    \Bigg\{\frac{2(1+\eta\nu)}{\beta} {a'_{\infty c} \over a_{\infty c}}
        {m'_c \over m_c}
        + {\eta\nu(2-\beta-\eta\nu) \over \beta^2}
                 \left({m'_c\over m_c}\right)^2\nonumber\\
  &&\qquad\qquad\qquad\quad + {a''_{\infty c} \over a_{\infty c}}
	 - {a'_{\infty c} \over a_{\infty c}} {k''_c \over k'_c}
      +\frac{\eta\nu}{\beta} \left( {m'_c\over m_c}
        {k''_c \over k'_c} - {m''_c\over m_c}\right)\Bigg\},
\end{eqnarray}
where the prime denotes differentiation and for brevity we have used
$a'_{\infty c} \equiv a'_\infty(\thc)$, etc., and so on.

The required scaling form for $W(M;T,\hinf)$ must be somewhat
more elaborate because
of the additional dependence on $\hinf$: see the original definition
(\ref{eq4.2:Wdef2}).  (Indeed, the result presented in
\cite{Fisher90b} is somewhat misleading since the dependence on $\hinf$ was
suppressed and the expressions given apply only for $\hinf=0$.)  Let us consider,
first, the parametric representation for $W$ following from~(\ref{eq4.2:Wrep})
with the aid of~(\ref{eq4.1:ntheta}).  Note that
specification of $T$ and $\hinf$ implies, via (\ref{eq4.1:parametric}),
parametric coordinates $\rinf$ and $\tinf$
(for the corresponding bulk phase) while the variation of $M$ at constant
$T$ and $\hinf$ can be described by coordinates $r$ and $\t$.  Thus
from~(\ref{eq4.2:Wrep}) and (\ref{eq4.1:parametric}) we obtain
\begin{equation}
W(r,\t;\rinf,\tinf) \approx r^{2-\alpha} n(\t) - \rinf^{2-\alpha}
n(\tinf) - \rinf^\Delta l(\tinf) [r^\beta m(\t) - \rinf^\beta m(\tinf)].
\end{equation}
When $t \ne 0$, the relation $t = r k(\t) = \rinf k(\tinf)$ leads to the
desired form
\begin{equation}
W(M;T,\hinf) \approx r^{2-\alpha} w(\t;\tinf),
\label{eq4.4:Wpar}
\end{equation}
where, with $\hinf = \rinf^\Delta l(\tinf)$ and (\ref{eq4.1:parametric}) for $M$, we
have
\begin{eqnarray}
w(\t;\tinf) &=& n(\t) 
 - l(\tinf) m(\t) \left| k(\t) / k(\tinf) \right|^\Delta\nonumber\\
 &&\qquad -[n(\tinf)-l(\tinf)m(\tinf)]
    \left| k(\t) / k(\tinf) \right|^{2-\alpha},\label{eq4.4:w}
\end{eqnarray}
while the singular part of the Helmholtz
free energy \textit{extended} \textit{into} \textit{the} \textit{two-phase}
\textit{region}, as discussed above, is given by (\ref{eq4.1:As})~\cite{npm}.

When $T=T_c$ we have $\t = \tinf = \thc$ and
$w(\t;\tinf)$ simplifies to yield
\begin{eqnarray}
w_c &\equiv& w(\thc;\thc) = \beta m_c l_c / (2-\alpha),\\
w'_c &\equiv& (dw/d\t)_{\t = \tinf = \thc} = [\beta m_c l'_c - (1-\beta) l_c
	m'_c] / (1 - \alpha).
\end{eqnarray}
Consequently $w(\t;\tinf)$ is continuous and smooth through $T=T_c(g)$ (or $\t =
\tinf = \thc$).  However, the presence of the powers of $\Delta$ and
$(2-\alpha)$ in~(\ref{eq4.4:w}) shows that the curvature $(d^2w/d\t^2)$ is
\textit{not} continuous through $T = T_c(g)$.  Nevertheless, this lack of
analyticity of $w(\t;\tinf)$ is not, of itself, expected to lead to
corresponding nonanalytic behavior in surface free energies, etc., since
the underlying bulk free energies and correlation lengths do vary
analytically.

Now we can express $W$ in the alternative scaling form
\begin{equation}
W(M;T,\hinf) \approx |M|^{\delta+1} Y_\pm(\mt; {\tilde{h}}_\infty) 
\end{equation}
where, with $A^\infty(\tilde{h}_\infty) = -n(\tinf)/|k(\tinf)|^{2-\alpha}$,
we have
\begin{eqnarray}
Y_\pm(\mt;\tilde{h}_\infty)
 &=& A^\infty(\tilde{h}_\infty)|B\mt|^{-\delta-1} - \hinf[M-\Minf(T,\hinf)]|M|^{-\delta-1}\nonumber\\
  &&\quad + \Y0\left[1
	+ \sum^\infty_{n=1} \Y{n} \left(\pm |\mt|^{-1/\beta}\right)^n\right].
\end{eqnarray}
[This 
reduces to the Fisher-Upton expressions when $\hinf \to 0$ for $\tilde t
> 0$ and $\tilde t < 0$ although Eq.~(7) of~\cite{Fisher90b}, should be
corrected by changing $(\pm|y|)^{-n/\beta}$ to read
$(\pm |y|^{-1/\beta})^n$.]\enspace  On rearranging for $|\mt| \to \infty$
we have
\begin{eqnarray}
Y_\pm(\mt;\tilde{h}_\infty)
&=& \Y0 \Big[1 \pm \Y1 |\mt|^{-1/\beta} -\hbox{sgn}(M) \Y\Delta |\mt|^{-\Delta/\beta}
	+ \Y{2-\alpha} |\mt|^{-(2-\alpha)/\beta}\nonumber\\
&&\qquad\qquad + \Y2 |\mt|^{-2/\beta} \pm \Y3 |\mt|^{-3/\beta} + \cdots \Big],
\label{eq4.4:Ymexp}
\end{eqnarray}
where the various coefficients are given explicitly by
\begin{eqnarray}
  \Y\Delta &\equiv& \hinf / B^\delta \Y0 |t|^\Delta,\\
  \Y{2-\alpha} &\equiv& \left[ A^\infty(\tilde{h}_\infty)
	 + (B^2/C^+)\tilde{h}_\infty\mt_\infty(\tilde{h}_\infty)\right] / 
	 B^{\delta+1}\Y0,\\
\Y0 &=& n_c / (m_c)^{\delta+1},\\
\Y1 &=& \frac{[m_c/B]^{1/\beta}}{k'_c}
    \left\{\frac{n'_c}{n_c} - (\delta+1)
	\frac{m'_c}{m_c}\right\},\\
\Y2 &=& \frac{[m_c/B]^{2/\beta}}{2[k'_c]^2}
    \Bigg\{-\frac{2(1-\alpha)}{\beta} \frac{n'_c}{n_c}
	\frac{m'_c}{m_c}
     + {(\beta-\alpha)(1+\delta) \over \beta}
	 \left({m'_c\over m_c}\right)^2\nonumber\\
  &&\qquad  + {n''_c \over n_c} - {n'_c \over n_c}
	{k''_c \over k'_c}
     + (1+\delta) \left( {m'_c\over m_c}
	{k''_c \over k'_c} - {m''_c\over m_c}\right)\Bigg\},
\end{eqnarray}
where $m''_c = m''(\thc)$, etc.

\subsection{Representation of the Noncritical Phase}
The scaling forms just discussed provide a satisfactory
representation of the phases~$\beta$, $\gamma$ and $\beta\gamma$ (see
Fig.~\ref{fig4:phase}) in the vicinity of the critical line and critical endpoint, but they
seem  to give no account at all of the noncritical or spectator phase,
$\alpha$.  To overcome this draw-back, Fisher
and Upton \cite{Fisher90a, Fisher90b} advanced a hypothesis, $\bf \Omega$,
which asserts that as regards the singular contributions near a critical
endpoint, the noncritical phase $\alpha$ (typically a vapor when $\beta$ and
$\gamma$ are liquids) can be replaced by a rigid, inert wall, say $\omega$,
characterized only by a nonzero surface field $h_1$ favoring, say, the bulk
critical phase $\beta$.

If one accepts the hypothesis $\bf \Omega$, one must consider a wall at,
say, $z=0$ with a corresponding wall free energy as introduced in (\ref{eq4.3:DF})
which, following a Landau
approach~\cite{Nakanishi}, may be expanded as
\begin{equation}
f_1(M_1) = -h_1 M_1 + c_2 M_1^2 + \cdots. \label{eq4.4:h1}
\end{equation}
The boundary condition~(\ref{eq4.3:M1}) determining
$M_1$ can be rewritten for the EdGF theory using~(\ref{eq4.3:edgf}),
(\ref{eq4.3:curlyg}), and~(\ref{eq4.3:edgfprof}) as
\begin{equation}
\D{f_1}{M_1} = 2 \left[W \thinspace \xi^2/2\chi \right]^{1/2}_{M=M_1}.
\label{eq4.4:df1}
\end{equation}

If the surface field near bulk criticality scales as $h_1 \sim
|t|^{\Delta_1}$ when $h_1 \to 0$, this relation leads to $\Delta_1 =
{1\over2}(2-\alpha-\eta\nu) = \mu-\beta$.  However, it is known that the
critical exponents for surface quantities such as $h_1$ and $M_1$ are
characterized by exponents $\Delta_1$, $\beta_1$, etc., that \textit{cannot}
be derived simply from the bulk exponents \cite{Binder, Diehl86, Diehl97}.
Hence the
exponent relation implied by~(\ref{eq4.4:h1}) for small $h_1$ is not, in
fact, valid.
However, this failure of the local functional theory is of no concern in
the present situation since we wish to keep the field $h_1$ \textit{fixed} \textit{and} \textit{nonzero}.  Hence the correct scaling combination $h_1/|\tilde
t|^{\Delta_1}$ \textit{diverges} when $\tilde t \to 0$ and hence $h_1$ does
not appear explicitly in the asymptotic scaling forms.  From our viewpoint,
then, it is only necessary to investigate \textit{large} $|h_1|$ and,
correspondingly, large $|M_1|$.  In a more general formulation
the $\alpha$ phase would be represented by an extra
(``third'') minimum in the extended free-energy function $A^\dagger(M,t,g)$
at some value $M=M_\alpha(t,g)$ with, say, $M_\alpha < 0$ corresponding to a
vapor phase: a fuller investigation \cite{Fisher90a, Fisher90b}
then shows that the hypothesis $\bf \Omega$ is justified within EdGF
theory.  Of course, the effective surface order $M_1$ does not become
indefinitely large in this formulation: however, for the singular behavior
near the critical endpoint we may imagine taking $M_1$ to $-\infty$, the
negative sign being chosen so that, as indicated, the density
$\rho_\alpha$ of the vapor phase lies below the critical density $\rho_c$
(corresponding to $M=0$) so that the wall field $h_1$ should induce a negative value of $M$.
This limit
entails, as will now be discussed, the subtraction of appropriate leading terms
that would otherwise appear as divergencies but which, in reality,
contribute only to noncritical `background' terms.

\subsection{Implications of an Unbounded Surface Field}
To use the integral expression~(\ref{eq4.4:Sigma}) for the surface tension
in the limit where $h_1$ and, hence, $M_1$ remain finite so that the scaled
variable $\mt \sim M/|t|^\beta$ becomes large near the wall at $z=0$ when
$t\to0$, we need to examine the integrand.  When $|\mt| \to \infty$ we find
\begin{eqnarray}
\left[W(M)\frac{\xi^2(M)}{2\chi(M)}\right]^{1/2}
    &=& |M|^{(\mu-\beta)/\beta} \left(\Y0\Z0\right)^{1/2} \Bigg\{1
	+ {(\Y1+\Z1)t \over 2|M/B|^{1/\beta}}\nonumber\\
    &&\quad -\hbox{sgn}(M) {\hinf/B^\delta \Y0 \over 2|M/B|^\delta}
	+ {\Y{2-\alpha}|t|^{2-\alpha} \over 2|M/B|^{\delta+1}}\nonumber\\
    &&\quad + {[4(\Y2+\Z2)-(\Y1-\Z1)^2] t^2 \over 8|M/B|^{2/\beta}}\nonumber\\
    &&\quad -\hbox{sgn}(M)
	 {(\Z1-\Y1)\hinf t/B^\delta\Y0 \over 4|M/B|^{1/\beta+\delta}}\nonumber\\
    &&\quad +{(\Z1-\Y1)\Y{2-\alpha}t|t|^{2-\alpha}\over4|M/B|^{1/\beta+\delta+1}}
    	+ {\cal O}\left(|\mt|^{-3/\beta}\right) \Bigg\}.
\label{eq4.4:integndexp}
\end{eqnarray}
On integration the leading term $|M|^{(\mu-\beta)/\beta}$
yields $|M_1|^{\mu/\beta}$ which diverges when
$M_1 \to -\infty$ since $\mu/\beta>0$.  Similarly, the second term yields a
divergence when $(\mu-1)/\beta > 0$, which applies when $d>2$.  The third term
gives $|M_1|^{(\mu-\Delta)/\beta}$ which vanishes in the limit
$M_1\to-\infty$ since $\mu<\Delta$ for $d=3$.  However, when $d=4$ it
diverges as $\ln |M_1|$.  Higher order terms do not yield
any divergencies when $M_1\to-\infty$.  Hence, to remove the divergence in the
integral when $M_1 \to -\infty$ for $2 \le d \le 4$, the first three terms
must be subtracted.

When $\Minf>0$ the integral in (\ref{eq4.4:Sigma}) runs through $M=0$ which is problematical
because the piece to be subtracted is singular at $M=0$.  To
avoid this, we split the integral at an arbitrary point $M^*<0$ and
make the subtraction only in the integral from $M_1$ to $M^*$: of course,
the values of $M^*$ should not affect the final results.
Finally, the subtracted form of the surface tension expression (\ref{eq4.4:Sigma})
can be written
\begin{equation}
\Sigma(t,\hinf;M_1) = I_1 + I_2 + I_3(M^*) - I_3(M_1) + f_1(M_1), 
\end{equation}
where the various contributions follow from
\begin{eqnarray}
I_1 &=& 2\int_{M^*<0}^{\Minf} dM\thinspace \left[ W(M) \xi^2(M)/2\chi(M)
	\right]^{1/2},\\
I_2 &=& 2\int_{M_1}^{M^*} dM\thinspace \Bigg[
	\left\{W(M) \xi^2(M)/2\chi(M)\right\}^{1/2}
    - \left(\Y0\Z0\right)^{1/2} |M|^{\mu/\beta-1}\times \nonumber\\
    &&\qquad\qquad\qquad
	\left\{1 + {(\Y1+\Z1)t \over 2|M/B|^{1/\beta}}
		+ {\hinf / B^\delta \Y0 \over 2|M/B|^\delta}
	\right\}\Bigg],\\
I_3(M_1) &=& 2\left(\Y0\Z0\right)^{1/2} \int^{M_1} dM\thinspace
    |M|^{\mu/\beta-1}\cr
    &&\qquad\qquad\qquad\qquad
	\times\left[1 + {(\Y1+\Z1)t \over 2|M/B|^{1/\beta}}
	+ {\hinf/B^\delta\Y0 \over 2|M/B|^\delta}\right].
\end{eqnarray}
The third, \textit{indefinite} integral, which together with $f_1(M_1)$ contains
all divergences in the limit $M_1\to-\infty$, can be performed analytically
yielding
\begin{eqnarray}
I_3(M) &=& -2\left(\Y0\Z0\right)^{1/2} \Bigg[
    {(-M)^{\mu/\beta} \over \mu/\beta}
    + {(\Y1+\Z1)B^{1/\beta}t \over 2(\mu-1)/\beta} (-M)^{(\mu-1)/\beta}\nonumber\\
  &&\qquad\qquad\qquad\qquad + \thinspace{\hinf \over 2 \Y0} \negthinspace \times \negthinspace
    \left\{\begin{array}{ll}\beta (-M)^{\mu/\beta-\delta}/(\mu-\Delta), &\quad d<4\\
       \ln [(-M)/B\rinf^\beta], &\quad d=4\end{array}
    \right\}
  \Bigg],\label{eq4.4:I3inf}
\end{eqnarray}
where for $d=4$ we have made the argument of the logarithm
dimensionless and scale-free; this is harmless because the extra
$\ln (B\rinf^\beta)$
term amounts merely to an additive constant.

\subsection{Singular Part of the Surface Tension}
Following the discussion, we now identify the finite, singular part of the
surface tension as
\begin{equation}
\Delta\Sigma(t,\hinf)
  = I_1 + I_2 + I_3^* \qquad \hbox{with} \qquad I_3^* \equiv I_3(M^*),
\label{eq4.4:divfree2}
\end{equation}
where the limit $M_1 \to \infty$ in any remaining $M_1$-dependence is to be
understood.
Note, furthermore that when $d < 4$, the diverging parts that have been
subtracted vary only linearly with $t$.  As mentioned, they
may thus be regarded as a part of the common analytic
background, $\Sigma_0(T)$, for the surface tension: see~(\ref{eq4.1:dsabr})
and~(\ref{eq4.1:dsab}).  In identifying $\Delta\Sigma$ for $d=4$
there is some unavoidable arbitrariness associated with the introduction of
the $\ln(B\rinf^\beta)$ term: but this is of little significance.

Finally, to employ
the parametric representations, we change the integration
variable from $M$ to $\t$ in the integrals $I_1$ and $I_2$.
From (\ref{eq4.1:parametric}), we have
\begin{equation}
\left(\PD{M}{\t}\right)_t = \hbox{\rm sgn}(t){k(\t)m'(\t) - \beta k'(\t)m(\t)
   \over |k(\t)|^{\beta+1}}|t|^\beta
\end{equation}
and on defining $\t^* < 0$ via
\begin{equation}
M^*/|t|^\beta = m(\t^*) / |k(\t^*)|^\beta
\end{equation}
we find
\begin{eqnarray}
I_1 &=& 2\thinspace\hbox{\rm sgn}(t)|t|^\mu \int_{\t^*}^{\tinf} d\theta\thinspace
  {k(\t)m'(\t) - \beta k'(\t)m(\t) \over |k(\t)|^{\mu+1}}
  \sqrt{a_\infty(\t)w(\t)},\label{eq4.4:I1par}\\
I_2 &=& 2\thinspace\hbox{\rm sgn}(t)|t|^\mu \int_{-\t_c}^{\t^*} d\theta\thinspace
  {k(\t)m'(\t) - \beta k'(\t)m(\t) \over |k(\t)|^{\mu+1}}
  \Bigg[ \sqrt{a_\infty(\t)w(\t)}\nonumber\\
  &&\qquad -\thinspace (\Y0\Z0)^{1/2} |m(\t)|^{\mu/\beta-1} \Bigg\{1
     + {(\Y1+\Z1)k(\t) \over 2|m(\t)/B|^{1/\beta}}
     + {\Y\Delta |k(\t)|^\Delta \over 2|m(\t)/B|^\delta}\Bigg\}\Bigg],
\label{eq4.4:I2par}
\end{eqnarray}
where $w(\t)$ is given by~(\ref{eq4.4:w}) (for fixed $\theta_\infty$),
while $I_3(M^*)$ can be written as
\begin{eqnarray}
I_3^* &=& -2\left(\Y0\Z0\right)^{1/2} |t|^\mu B^{\mu/\beta} (y^*)^{-\mu} \Bigg[
    {\beta\over\mu}
    + \hbox{sgn}(t) {(\Y1+\Z1)y^* \over 2(\mu-1)/\beta}\qquad\qquad\qquad\thinspace \nonumber\\
  &&\qquad\qquad\qquad + \thinspace {\Y\Delta\over2}
    \left\{\begin{array}{ll}\beta (y^*)^\Delta/(\mu-\Delta), &\quad d<4\\
       \beta (y^*)^\mu\ln (|k_\infty| / y^*), &\quad d=4\end{array}
    \right\}
  \Bigg],\label{eq4.4:I3par}
\end{eqnarray}
where $y^* = |k(\t^*)| [-m(\t^*)/B]^{-1/\beta}$.  It is to be understood that,
below $T_c$, the integral $I_1$ must be interpreted as $\int^{\tinf}_{\t^*} =
\int^{-\t_0}_{\t^*} + \int^{\tinf}_{\t_0}$ since, by our periodic
parametric construction, $\t = +\t_0$ and $\t = -\t_0$ are identified
together with $M = 0$.

These expressions for $I_1$, $I_2$, and $I_3^*$ together with
(\ref{eq4.4:divfree2}) for $\Delta\Sigma(t,\hinf)$ constitute our explicit
results for the singular part of the surface tension.
Note that $M_1$, the boundary value of the
profile on the wall that stands in for the $\alpha$ phase, does \textit{not}
enter these expressions.  On the other hand the arbitrary value $\t^*$
does appear; however, the total $I_1 + I_2 + I_3^*$ should be independent of
$\t^*$ and this may be checked in explicit calculations.

\section{Parametric Scaling Function for $d=3$}
In this section we examine the numerical results for the parametric scaling
function $s(\t)$ introduced in (\ref{eq4.1:sigma}) that follow from the EdGF surface
tension expressions obtained in the previous section, specifically (\ref{eq4.4:divfree2})
with (\ref{eq4.4:I1par})--(\ref{eq4.4:I3par}).

\subsection{Numerical Evaluation of the Scaled Tension}
Now given the parametric angular functions $k(\t)$, $m(\t)$, etc.,
the numerical integration of $I_1$ in (\ref{eq4.4:I1par}) can be performed readily.
However, there is an
endpoint singularity in $I_2$ in (\ref{eq4.4:I2par}) at $\t\negthinspace = \negthinspace-\t_c$ where $k(\t)$ vanishes.
On using
the expansion~(\ref{eq4.4:integndexp}) for $[W\xi^2/2\chi]^{1/2}$
one sees that the singularity has the form
$|\t-\t_c|^\phi$ with $\phi = 1-\alpha-\mu = \nu-1 < 0$ for $d = 3$.
Although divergent, this singularity
is integrable and can be handled by standard techniques~\cite{Press}.

As a check on the numerical calculations, it is useful to recall
the mean-field limit
for which exact analytic results can be obtained.  Indeed, from the 
asymptotic classical
equation of state, namely,
\begin{equation}
h(M,t) = D M(B^2 t + M^2) \quad\hbox{with}\quad D=1/B^2C^+,
\label{eq4.4:mfeq}
\end{equation}
one can find the auxiliary free energy via~(\ref{eq4.2:Wrep})
to obtain
\begin{equation}
W(M) = \hbox{$1\over2$} t B^2 D (M^2-\Minf^2) + \hbox{$1\over4$}D(M^4-\Minf^4)
	- \hinf(M-\Minf),
\end{equation}
in which $\Minf \negthinspace = \negthinspace M(T,\hinf)$ is,
of course, the appropriate bulk equilibrium
value.
Using the EdGF formulation and taking $\xi^2/2\chi = {1\over2} J_0$ [see
(\ref{eq4.2:mfDf})],
the critical surface tension can be calculated analytically, yielding
\begin{eqnarray}
\Sbr &=& 2 \int^{M_0(T)}_{-M_0(T)} dM\thinspace
    \left[W(M)\xi^2/2\chi\right]^{1/2} \\
 &=& \hbox{$2 \over 3$} \sqrt{2 J_0 / B^2 C^+} M_0^3(T),\nonumber
\end{eqnarray}
from which the amplitude $K$ defined in~(\ref{eq4.1:dsbr}) is
\begin{equation}
K = \hbox{$2 \over 3$} K_0 \qquad \hbox{with} \qquad K_0 = \sqrt{2 J_0 B^4 / C^+}.
\label{eq4.4:mfK}
\end{equation}
Similarly, the amplitudes $K^\pm$ for the noncritical surface tension can be
calculated using the subtraction scheme
discussed above; they are found to be
\begin{equation}
K^+ = -\hbox{$1\over3$} \sqrt{2} K_0, \quad
K^- = \hbox{$1\over3$} K_0,
\label{eq4.4:mfKs}
\end{equation}
which values confirm the
universal amplitude ratios $P$ and $Q$ stated in~(\ref{eq4.1:PQmf}).

For the numerical results presented below we have used
the extended sine model expounded in~\cite{Zinn99},
specifically in Eqns.~(5.6) and (7.4) with (6.1) and (4.4).
The numerical parameter values for $d=3$ are given in Eqs.~(6.2) and~(7.5) of
Ref.~9. 
In light of more recent estimates for the critical exponents and universal
amplitude ratios of $d\negthinspace=\negthinspace3$ Ising models~\cite{Compostrini2002, Butera2000, Engels2003},
the model parameters should, ideally, be updated.  However, the resulting
changes could induce only
rather small effects.  Thus the new estimates $\nu \simeq 0.6302$ and $\gamma \simeq 1.2375$~\cite{Butera2000}
differ from the adopted values of~\cite{Zinn99} by only 0.16\% and 0.28\%,
respectively.  The universal amplitude ratio estimates $f^+_1/f^-_1 = 1.963 \pm 8$ and
$C^+/C^- = 4.762 \pm 8$~\cite{Butera2000} agree well with those used in
\cite{Zinn99} lying within the uncertainty ranges although their central values are shifted
by 0.15\% and 3.8\%, respectively.  Furthermore, it has been shown via
parameter-sensitivity checks that the associated variations in universal ratios are less
than 5\% \cite{Zinn99}.  Thus the accepted parameter values from~\cite{Zinn99} are perfectly reasonable
for the numerical aspects of the calculations reported here.

The computation may be set up to generate the
mean-field results applicable when $d>4$: see Appendix~D of Ref.~17 
for the appropriate parameter values.
Our numerical values for the mean-field case agree up to eight digits and the lack of any dependence on
$\t^*$ was verified.  (For convenience the values $\t^* = -3\t_c/4$ for $0\le\tinf<\t_c$ and
$\t^* = -(\t_c+\t_0)/2$ for $\t_c < \tinf \le
\t_0$ were adopted \cite{Zinn97}.)
In $d=3$ dimensions the value of $\t^*$ was varied from $-0.05$ to
$-0.26$ for a sample value of $\tinf$ near $\t_c$: no change in $s(\t)$ was found
within eight-digit precision.

\subsection{Calculated Scaling Functions}
\begin{figure}
\includegraphics{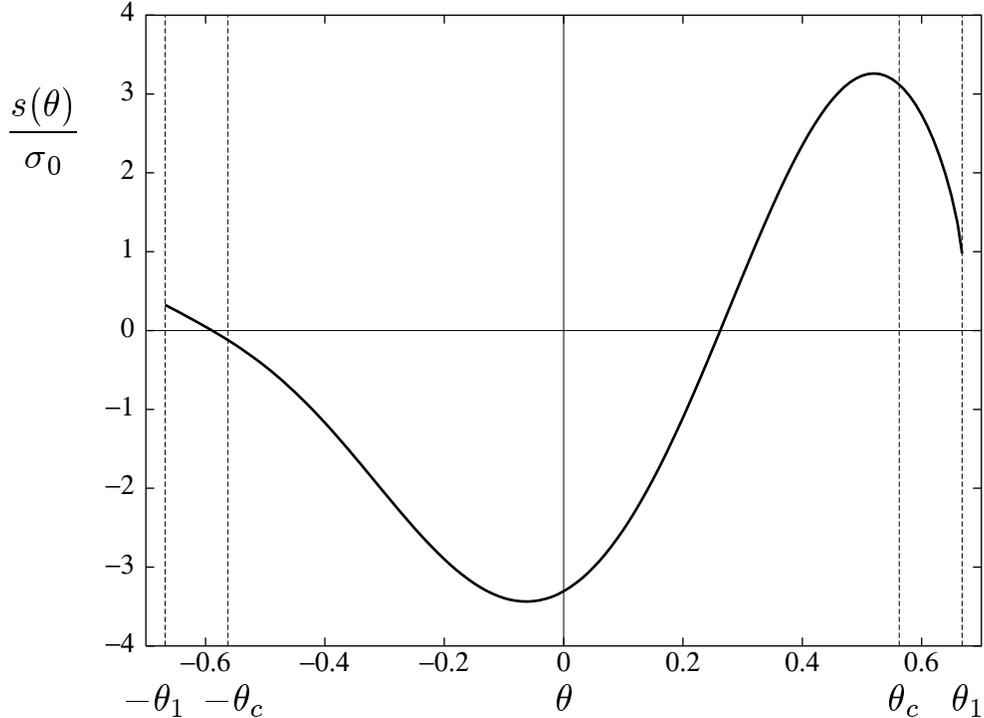}%
\caption{\label{fig4:s4} The angular scaling function
$s(\theta)$ for the surface tension in $d = 4$ dimensions
as given by EdGF theory.  Some significant numerical
values are $\theta_c \simeq 0.562\thinspace345$ and $\theta_1
\simeq 0.667\thinspace708$ corresponding to $s_c/\sigma_0=3.120\thinspace23$, $s_{-c}/\sigma_0
=-0.122\thinspace322$, $s_1/\sigma_0=0.976\thinspace225$, and $s_{-1}/\sigma_0=0.325\thinspace408$.
Note also $s(0)/\sigma_0 = -3.306\thinspace20$ \cite{Zinn97}.}
\end{figure}

Our calculations may be summarized by presenting the angular
function $s(\t)$ for the surface tension.
For the purpose of normalization, we employ
\begin{equation}
\sigma_0 = (a_{\infty} l_0 m_0^3)^{1/2},
\label{eq4.4:sigma0}
\end{equation}
which has the same units as surface tension.
Fig.~\ref{fig4:s4} presents a plot of $s(\t)$ for $d=4$.
One sees that, owing to the wall dependence (embodied in $M_1 \to -\infty$),
$s(\t)$ is \textit{not} symmetric with respect to $\t=0$.
Notice also that $s(\t)$ varies perfectly smoothly across the
critical isotherm values $\t=\pm\t_c$ as it should.

\begin{figure}
\includegraphics{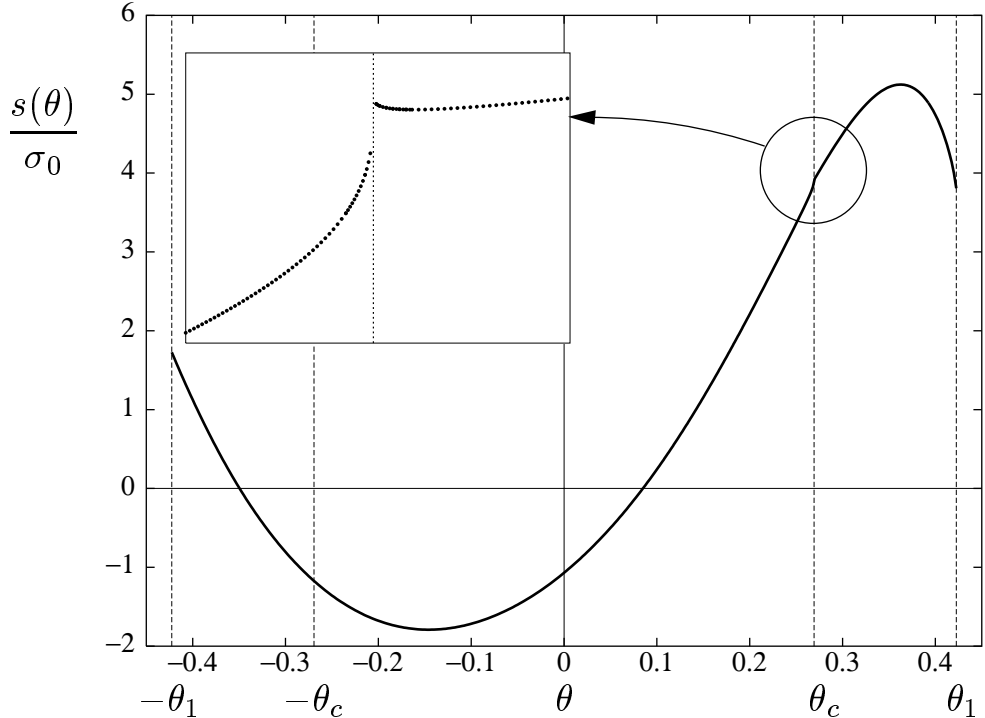}
\caption{\label{fig4:s3} The calculated angular function for
the surface tension in $d=3$ dimensions based on EdGF theory
and the extended sine model~\cite{Zinn99}.
The inset clearly shows a cusp at $\theta = \theta_c$
(i.e.~at $T=T_c$ for $\hinf > 0$) with a corresponding value $s_c/\sigma_0
= 3.917\thinspace31$.  The values of $s_{-c}$ and $s_{\pm1}$ are entered in Table~I.}
\end{figure}

The qualitative features of the scaling function in $d=3$ dimensions
are the same as for $d=4$.  However, as seen in Fig.~\ref{fig4:s3},
an unexpected, albeit small cusp appears at $\t=\t_c$.
The presence of the cusp (which is
certainly absent when $d=4$) implies that there is a line of nonanalyticity
along the
critical isotherm $\tilde t = 0$ when $\hinf>0$.  This is quite unphysical since
singularities in the thermodynamic functions, including surface quantities,
can occur only at the critical point $(t,\hinf) = (0,0)$.  Thus, from the
analyticity away from the critical point, one expects
\begin{equation}
\Delta\Sigma(t,\hinf) = \Delta\Sigma_c(\hinf)
	 + \Delta\Sigma_1^\pm(\hinf) t + \Delta\Sigma_2^\pm(\hinf) t^2 + \cdots,
\label{eq4.4:DSc}
\end{equation}
when $\hinf \ne 0$ and $t \to 0\pm$, respectively.  However,
analysis shows that the surface tension
predicted by the EdGF theory behaves as
\begin{eqnarray}
\Delta\Sigma(t,\hinf) &=& \Delta\Sigma_c(\hinf)
	 + \Delta\Sigma_{1\over2}^\pm(\hinf) |t|^{\mu-1+\alpha/2}\nonumber\\
    &&\qquad+ \Delta\Sigma_1^\pm(\hinf) |t|^{\mu + 1 -\alpha/2 - \Delta} + \cdots,
    \quad \hinf > 0,\label{eq4.4:cusp}
\end{eqnarray}
when $t \to 0\pm$: see Appendix~A in Ref.~17. 
In $d\ge4$ one has $\mu - 1 + {1\over2} \alpha = {1\over2}$ so one should,
in principle, see a square root cusp then; but such a cusp is absent 
in Fig.~\ref{fig4:s4}.  This is because
the amplitudes $\Delta\Sigma_{1\over2}^\pm(\hinf)$ vanish
identically in the classical situation: see
Eq.~(A.30) of Ref.~17. 

It transpires, as we will now explain, that this erroneous singular
behavior appears because the EdGF theory does not satisfy the
desideratum~${\bf I}$
of Sec.~III.  More specifically, a zero of the order parameter
profile is not represented correctly when $T \simeq T_c$ and $\hinf > 0$: recall
that the order parameter in the EdGF surface tension integral
(\ref{eq4.4:Sigma}) now runs from $M=M_1<0$ to $M=\Minf>0$ and so passes
through the value $M=0$ which is critical when $T=T_c$ or $\t=\t_c$.
Equivalently, the EdGF surface
tension can be represented by a real-space integral involving the free
energy~$W[M(z)]$, as in~(\ref{eq4.3:edgfsig2}).  Now, let us focus
on the term $\hinf M(z)$ in $W(M)$ [see~(\ref{eq4.2:Wdef2})]; \ie, consider
the contribution of $\int\hskip -0.5 ex dz\thinspace \hinf M(z)$
to the surface tension.  In
order to find the scaling behavior of the profile near $M=0$, we may analyze
(\ref{eq4.3:edgfprof}) to find (see Appendix B of Ref.~17 
for details)
\begin{equation}
M(z) \propto t^{\eta\nu/2} (z-z_0),\label{eq4.4:Matz0}
\end{equation}
when $z \to z_0$.  Evidently the
$z$-dependence is linear as expected from the analyticity of $M(z)$
when $t \ne 0$.
However, by scaling $z$ varies as $\xi \sim t^{-\nu}$ and
$\hinf$ as $t^\Delta$; thus the scaling behavior of the profile $M(z)$
near $z=z_0$ can be estimated leading to
\begin{equation}
\int dz \thinspace \hinf M(z) \sim
  t^{\eta\nu/2} \xi^2 \hinf \int {dz\over\xi} {(z-z_0)\over \xi}
  \sim t^{(\eta\nu/2)-2\nu+\Delta}.
\end{equation}
Note that the second integral is dimensionless and scale-free
and so has been regarded as a fixed number.
Via the scaling relations, the exponent
${(\eta\nu/2)-2\nu+\Delta}$ reduces to ${\mu-1+(\alpha/2)}$ which
corresponds to the cusp
appearing in~(\ref{eq4.4:cusp}).

Since the correct scaling function should not exhibit such nonanalyticity,
we introduce an interpolation procedure that smooths out
the calculated cusp in $s(\t)$.  The $\t$ range, say $[\t_a, \t_b]$, selected for
the interpolation is arbitrary.  However, the choice $(\t_a, \t_b)
= (0.20, 0.35)$ encompasses the cusp and
seems reasonable.  We have used a polynomial $P(\t)$
of degree 5 which is the minimal
order to match the first and second derivatives at both
ends of this interval.  The resulting polynomial for the extended sine model
parameters is given in Eq.~(4.4.50) of \cite{Zinn97}.  It agrees closely
with the calculated values of $s(\t)$ in 
the intervals $[0.20, 0.26]$ and $[0.32, 0.35]$;
Furthermore, the
largest deviation of $P(\t)$ from $s(\t)$ occurs at the cusp and is
only 3\%.


From now on, when the distinction matters, we will write
${\overline s}(\t)$ for the ameliorated
angular function to distinguish it from $s(\t)$ that has the cusp at $\t =
\t_c$.
Selected numerical values of ${\overline s}(\t)$
are given in Table~I; values spaced at intervals $\Delta \t = 0.01$ are
available in Table 4.1 of \cite{Zinn97}.

\begin{table*}
\caption{Numerical values for the ameliorated angular
surface tension scaling function ${\overline
s}(\theta)$ in three dimensions.
Note that $\t_c = 0.269\thinspace293$ and $\t_1 = 0.422\thinspace519$~\cite{Zinn99}.}
\begin{ruledtabular}
\begin{tabular}{rrrrrrrr}
\hfil $\theta$ & \hfil${\overline s}(\theta)/\sigma_0$ &
\hfil $\theta$ & \hfil${\overline s}(\theta)/\sigma_0$ &
\hfil $\theta$ & \hfil${\overline s}(\theta)/\sigma_0$ &
\hfil $\theta$ & \hfil${\overline s}(\theta)/\sigma_0$\\
\hline
$-\t_1$ &\sign$1.727\thinspace13$	&$-0.20$ &$-1.677\thinspace43$	&\sign$0.08$ &$-0.076\thinspace19$	&$0.30$ &$4.458\thinspace74$\\
$-0.40$ &\sign$1.125\thinspace36$	&$-0.16$ &$-1.783\thinspace25$	&\sign$0.12$ &\sign$0.587\thinspace32$	&$0.32$ &$4.792\thinspace82$\\
$-0.36$ &\sign$0.207\thinspace79$	&$-0.12$ &$-1.767\thinspace07$	&$0.16$ &$1.355\thinspace90$	&$0.34$ &$5.022\thinspace93$\\
$-0.32$ &$-0.513\thinspace65$		&$-0.08$ &$-1.638\thinspace86$	&$0.20$ &$2.210\thinspace93$	&$0.36$ &$5.121\thinspace25$\\
$-0.28$ &$-1.055\thinspace67$		&$-0.04$ &$-1.405\thinspace58$	&$0.24$ &$3.129\thinspace71$	&$0.38$ &$5.049\thinspace63$\\
$-\t_c$ &$-1.173\thinspace07$		&\sign$0$ &$-1.069\thinspace44$	&\hfil$+\t_c$ &$3.812\thinspace64$	&$0.40$ &$4.739\thinspace23$\\
$-0.24$ &$-1.438\thinspace71$		&\sign$0.04$ &$-0.627\thinspace87$&$0.28$ &$4.050\thinspace87$	&\hfil$+\t_1$ &$3.806\thinspace99$\\
\end{tabular}
\end{ruledtabular}
\end{table*}

In the previous study of the extended sine model~\cite{Zinn99},
various parameter sets near the preferred set in Eq.~(6.2) of~\cite{Zinn99}
were examined in order to
check the sensitivity of the universal amplitude ratios.  Here, we also check
the sensitivity of $s(\t)$ under the variations in the parameter sets
considered in Table~I of~\cite{Zinn99}.
As explained in Sec.~VII of \cite{Zinn99}, the optimal
parameters $a_{\infty 0}$, $a_{\infty 2}$, etc., for the true
correlation length are determined separately for each set by
fitting the universal ratios
$\thinspace\thinspace K(f^-)^2$ and
$\thinspace\thinspace \alpha A^+ \left(f^+\right)^d$, $\thinspace\thinspace f^+/f^-$,
and $\thinspace\thinspace \left(C^+/C^c\right) \left(f^c/f^+\right)^{2-\eta}$.
Then $s(\t)$ is computed for each parameter set using EdGF theory.

It must be noted that the value of $\t_1$ depends somewhat on the parameter
set.  This implies that the values of $s(\t)$
calculated from different parameter sets are not strictly comparable.
However, the change in $\t_1$ occurs only in the third
decimal place.  Thus, ignoring the small changes in the $\t$ scale, we have
examined $\Delta s(\t) \equiv s(\t) - s(\t)_{0}$, \ie,
the deviations of $s(\t)$ calculated using the sets $1, 2, \ldots, 6$ in
Table~I of~\cite{Zinn99} in place of
the preferred set~(6.2) of Ref.~9. 


The largest deviation occurs at $\t = \t_1 \simeq 0.42$; normalized by the
value ${\overline s}(\t_1)$ in Table~I it is about 3\%.  In~\cite{Zinn99}
the variations in the predicted universal ratios associated
with the parameter changes ranged from 0.2\% to 5\%.  Thus, the function
$s(\t)$ shows only the same level of variation as might have been
anticipated.

Finally, we have also checked the effects on $s(\t)$ of using ``untuned''
estimates for $a_\infty(\t)$.  Indeed it was found in
\cite{Zinn99} that the $[2/0]$ Pad\'e
approximant for $a_\infty(\t)$ could not fit
the universal ratio
$K(f^-)^2$ while also fitting $f^+/f^-$, $\alpha A^+ (f^+)^3$, and
$(C^+/C^c)(f^c/c^+)^{2-\eta}$.  Likewise for the other approximants of the same
order, namely, $[1/1]$ and $[0/2]$.
Hence, it was necessary to
introduce the $[3/0]$ approximant.
By using the $[2/0]$ approximant for $a_\infty(\t)$,
which is representative of the low-order approximants, we have
calculated the angular function $s(\t; [2/0])$ with
the preferred parameter set~(6.2) of~\cite{Zinn99}.
The difference between
$s(\t; [2/0])$ and $s(\t; [3/0])$ in the range $-\t_1 \le \t \le \t_c$ is
very small; indeed, it is invisible on a plot.  However, a large
difference of about 30\% arises in the subcritical range $\t_c \le \t \le \t_1$.
This occurs (i) because the approximants $[2/0]$ and $[3/0]$
behave quite differently in the two-phase region $\t_1 \le |\t| \le \t_0$
(see Fig.~8 of~\cite{Zinn99}) and (ii) because the calculation of
$s(\tinf)$ for $\t_c < \tinf < \t_1$ involves integration through the two-phase
region whereas the range $-\t_1 \le \tinf < \t_c$ requires
integration only through the one-phase region in which the approximants
are very similar.

\section{Numerical Results for the Surface Tensions}
Based on the angular scaling function $s(\t)$ or, where appropriate, the ameliorated version
${\overline s}(\t)$,
calculated with the preferred parameter set (6.2) and (7.5) of
Ref.~9, 
we now describe various theoretical predictions for the interfacial tension near a critical
endpoint.

\subsection{Amplitude Ratios}
Recall first the surface tension amplitudes $K^\pm$ and $K$
defined in~(\ref{eq4.1:dsbr})--(\ref{eq4.1:dsab}).  On using the
parametric form for $\Delta\Sigma$ in~(\ref{eq4.1:sigma})
the surface tension amplitude above $T_c$ is readily read off as
\begin{equation}
K^+ = s(0) / [k(0)]^\mu = s(0).
\end{equation}
Owing to the
asymmetry of ${\overline s}(\t)$ with respect to $\t \ge 0$, we must choose
the sign of $\pm\t_1$ appropriately for zero-field below $T_c$.  In accord with
Fig.~\ref{fig4:phase}(a), the $\alpha|\beta$
interface exists beneath $T_c$ when $h \le 0$; thus, the amplitude $K^-$
must be evaluated at $\t = -\t_1$
yielding
\begin{equation}
K^- = s(-\t_1) / |k(\t_1)|^\mu. \label{eq4.5:Km}
\end{equation}
Similarly, at $\t = \t_1$ one obtains the amplitude for $\Delta\Sar$ in
the limit $h \to 0+$, which via Antonow's rule is the sum $(K + K^-)$:
thus we have
\begin{equation}
K = [s(\t_1) - s(-\t_1)] / |k(\t_1)|^\mu.
\end{equation}

It is also instructive to define corresponding critical surface tension amplitudes
\textit{on} the critical isotherm.
Via scaling one can write
\begin{equation}
\Delta\Sigma \approx K^c_{\gl} |h|^{\mu/\Delta} \approx K^{\gl}_M |M|^{\mu/\beta},
\label{eq4.5:Kcpm}
\end{equation}
where the sub- and superscripts $>$ and $<$ stand for $h,M>0$ and $h,M<0$, respectively:
owing to the asymmetry of the surface tension with respect to $h \to
-h$, this distinction is essential.  For reference we may note that $\mu / \Delta
\simeq 0.80_6$ while $\mu / \beta = 3.8_5$.
The parametric representations then yield
\begin{equation}
K^c_> = {\overline s}(\t_c) / |l(\t_c)|^{\mu/\Delta}, \quad
K^c_< = s(-\t_c) / |l(\t_c)|^{\mu/\Delta}.
\end{equation}
Note the appearance of ${\overline s}(\t)$ here so that the prediction for
$K^c_>$ depends on the amelioration procedure.

The specific reduced surface tension amplitudes predicted by our EdGF
theory with the extended sine model are thence
\begin{eqnarray}
  &&K^+/\sigma_0 = -1.069\thinspace44, \quad K^-/\sigma_0 = 1.281\thinspace68,
	\quad K/\sigma_0=1.543\thinspace44,\nonumber\\
  &&K^c_>/\sigma_0 l_0^{-\mu/\Delta} = 0.293\thinspace755, 
	\quad K^c_</\sigma_0 l_0^{-\mu/\Delta} = -0.090\thinspace382\thinspace6,\nonumber\\
  &&K^>_M/\sigma_0 m_0^{-\mu/\beta} = 923.242,
	\quad K^<_M/\sigma_0 l_0^{-\mu/\beta} = -284.063,
	\label{eq4.5:Ks}
\end{eqnarray}
where we used Table~I for the ${\overline s}(\t)$ values and the extended
sine model values~\cite{Zinn99}
\begin{eqnarray}
&&k(\t_1) = -1.266\thinspace16, \quad l(\t_c)/l_0 = -l(-\t_c)/l_0 = 0.529\thinspace162,
\nonumber\\
&&m(\t_c)/m_0 = -m(-\t_c)/m_0 = 0.240\thinspace366,
\end{eqnarray}
while $\sigma_0$ is defined in (\ref{eq4.4:sigma0}) and represents the nonuniversal surface tension
scale.

From Eq.~(\ref{eq4.5:Ks}), one obtains the universal
amplitude ratios
\begin{equation}
P = 0.137_5 \pm 2, \quad Q = -0.834 \pm 2,\label{eq4.5:edgfPQ}
\end{equation}
where the uncertainties have been estimated by examining results from the other
parameter sets in Table~I of Ref.~9. 
In comparison with the preliminary calculations
quoted in~(\ref{eq4.1:PQfu}),
one sees that our improved estimate
for $P$ 
is about 15\% larger
[and still \textit{positive} in contrast to (\ref{eq4.1:PQmf})] while
the $Q$ value displays only a 0.5\%
deviation.   These estimates will be discussed elsewhere~\cite{FisherNote}
in relation to the experimental observations of Mainzer-Althof and
Woermann~\cite{Mainzer}.

For the surface tensions on the critical isotherm, our analysis generates the
universal amplitude ratio predictions
\begin{eqnarray}
&&{K^c_> \over K}\left( {B \over C^+} \right)^{\mu/\Delta} =
\hphantom{-}3.34_{\hphantom{9}}\pm4, \quad
 {K^>_M \over K} B^{\mu/\beta} = \hphantom{-}5.26\pm7,\nonumber\\
&&{K^c_< \over K}\left( {B \over C^+} \right)^{\mu/\Delta} = -1.02_9\pm1, \quad
 {K^<_M \over K} B^{\mu/\beta} = -1.62\pm2,
	\label{eq4.5:Kcpmratio}
\end{eqnarray}
where we recall that $B$ and $C^+$ are defined as in~\cite{Zinn98}.
More directly we find
\begin{equation}
K^c_> / K^c_< = K^>_M / K^<_M = -3.25 \pm 5,
\end{equation}
which may also be used in analyzing experimental data and might well be tested
in simulations of Ising-type systems.

We plan, as mentioned in the Introduction, to consider the applications of the
present theory to experiments in the future~\cite{FisherNote}; but it should
be noted here that the difference in sign and magnitude of the amplitudes
$K^>_M$ and $K^<_M$ was explicitly remarked by \textit{NWW} on the basis of
the theory of Ramos-G\'omez and Widom~\cite{Widom}, a square-gradient approach
formulated to incorporate $\delta = 5$.  [See also Rowlinson and
Widom~\cite{Rowlinson} (pp.~287--293) and our comments following
(\ref{eq4.1:Antonow}) and
(2.10) above.]  Furthermore, their analysis leads to $K^>_M / K^<_M \simeq
-148/42 \simeq -3.5_2$: see~\cite{Widom} p.~614 and~\cite{Rowlinson}
Eq.~(9.124).  This value is only some 8\% larger in magnitude than found here.

\subsection{Scaling Functions}
\begin{figure}
\includegraphics[scale=0.8]{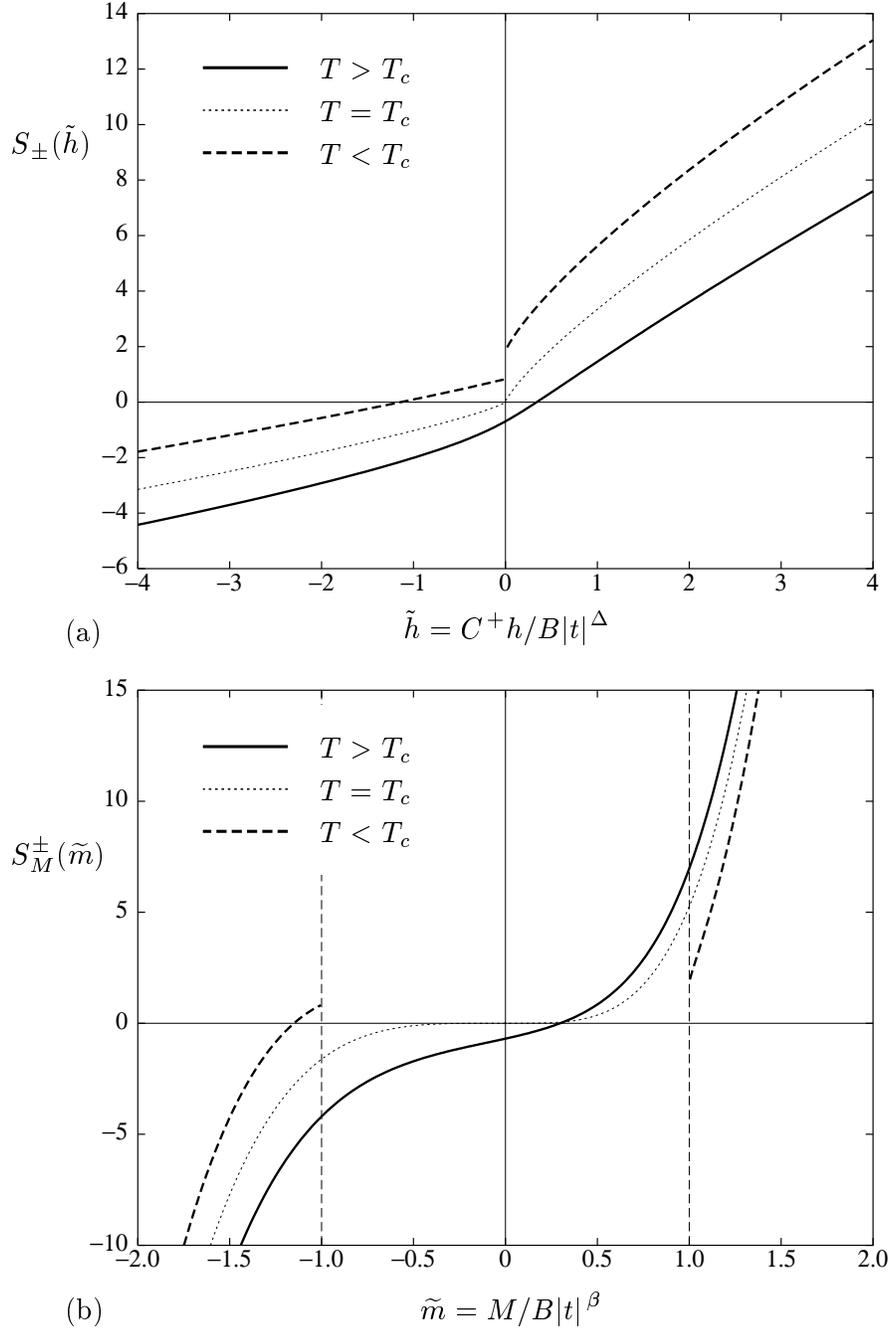}
\caption{\label{fig4:Spm}
The universal scaling functions predicted by the EdGF
surface tension theory (a) vs the ordering field $\tilde h$ and (b) vs
the order parameter $\widetilde m$.
Note the lack of symmetry about $\tilde h = \widetilde m = 0$
and the jumps dictated by (\ref{eq4.5:SMpmnorm}).}
\end{figure}

While parametric scaling forms are conceptually and computationally effective,
direct scaling representations, as in (\ref{eq4.1:sigma}), are more useful for
comparison with existing or proposed observations.  Thus from the angular
function ${\overline s}(\t)$ we have computed
the scaling functions $S_\pm(\tilde h)$: in Fig.~\ref{fig4:Spm}(a)
these are plotted in terms of the field (or chemical potential) variable
$\tilde h$.

In experiments, for example on binary mixtures, the density deviation
$(\propto M)$ is
more readily accessible than the chemical potential deviation $(\propto \tilde h)$.
Hence, in practice scaling plots in terms of $\widetilde m$ are
more convenient: the corresponding scaling functions may be defined via
\begin{equation}
\Delta\Sigma \approx K |t|^\mu S_M^\pm(\mt) \quad \hbox{with} \quad
\mt = M/B|t|^\beta,\label{eq4.5:SMpm}
\end{equation}
while their behavior is shown in Fig.~\ref{fig4:Spm}(b).
The normalizations adopted in (\ref{eq4.1:sigma}) and here lead, with Antonow's rule
(\ref{eq4.1:Antonow}) to the ``jump conditions''
\begin{equation}
S_-(0+) - S_-(0-) = S_M^-(1) - S_M^-(-1) = 1,
\label{eq4.5:SMpmnorm}
\end{equation}
that enter below $T_c$.

Away from the critical point, all bulk and surface quantities must be analytic
unless some phase boundary intervenes.  This is reflected in
the smooth behavior of $S_+(\tilde h)$ for all $\tilde h$.  In contrast,
there is a jump in $S_-(\tilde h)$ at $\tilde h=0$ in accordance with
(\ref{eq4.5:SMpmnorm}): for $\tilde h \ne 0$, however, $S_-(\tilde h)$
is analytic for all $\tilde h$.
Similar considerations apply to the
break in the $S^-_M(\mt)$ plot seen in Fig.~\ref{fig4:Spm}(b).
On the critical isotherm $T=T_c$
the two branches of the scaling function, $S_+(\tilde h)$ and
$S_-(\tilde h)$, must join
smoothly when $\tilde h \to \pm\infty$;
this again is a consequence
of the overall requirement of analyticity away from
the phase boundary.  Hence, the two
branches of $S_\pm(\tilde h)$ and $S^\pm_M(\mt)$ must both asymptotically approach one
another when $\tilde h \to \pm\infty$ and $\widetilde m \to \pm\infty$:
this can be seen easily in Fig.~\ref{fig4:Spm}(b).
Notice that the dotted line plots labelled $T = T_c$ in Fig.~\ref{fig4:Spm}
depict only the asymptotic power-law behavior embodied in (\ref{eq4.5:Kcpm}) that
corresponds in these scaled plots to $\tilde h \to \pm \infty$.
(It may also be
remarked that the unphysical cusp in the EdGF surface tension on the critical isotherm for
$h>0$ is located at $\tilde h \to +\infty$; thus, with or without the cusp,
the scaling plots in Fig.~\ref{fig4:Spm} look similar.)

Finally, it should be clear that the scaling function plots in Fig.~\ref{fig4:Spm} may also
be read as describing the variation of $\Delta \Sigma (T, h) \propto S_\pm(\tilde h)
= S^\pm_M(\widetilde m)$ with $h$ and $M$ at fixed values of $T \gl T_c$.
In particular, one may then notice that the isotherms of $\Delta \Sigma$ vs $M$ for
$T$ above $T_c$ will \textit{cross} the critical isotherm when $M$ is positive,
and, by continuity, hence also cross one another~\cite{FisherNote}.
This crossing of the surface tension isotherms above~$T_c$ has been anticipated
theoretically by Ramos-G\'omez and Widom~[10: see text after their Eq.~(3.15) and
Table~I] who also point to some experimental evidence of crossings.

\begin{figure}
\includegraphics[scale=0.8]{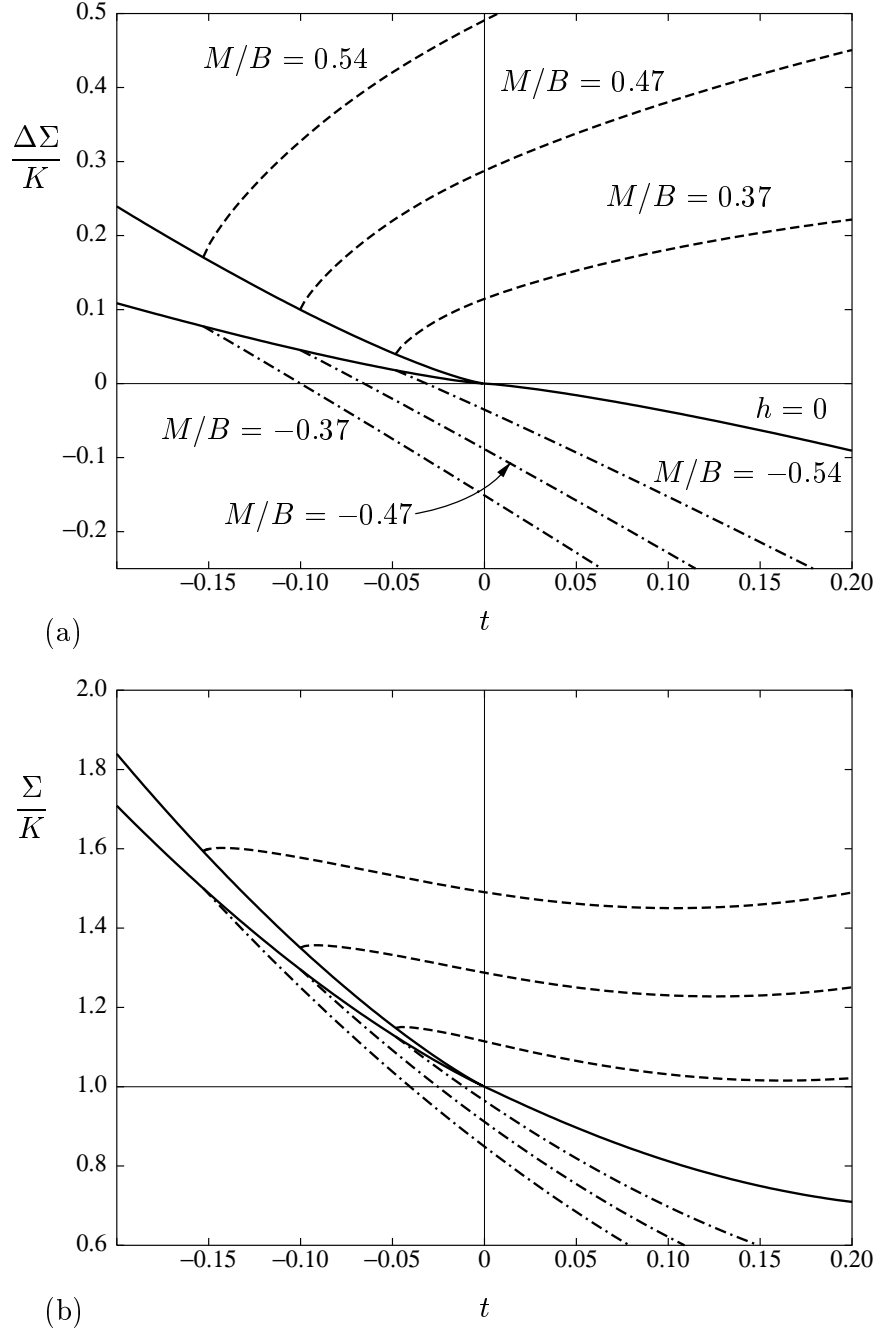}
\caption{\label{fig4:isochore}
Plots of (a) the singular part of the critical endpoint surface tension
as a function of the reduced temperature $t$ for fixed values of
$M$ and (b) with the addition of the model background term $\Sigma_0(T,h)$ given
in~(\ref{eq4.5:S0}) for the same values of $M$.}
\end{figure}

On the other hand the isotherms of $\Delta \Sigma$ vs $[h - h_0(T,g)]$
(which, recalling (\ref{eq4.4:norm})
et seq., should replace $h$ in the figure) do \textit{not} cross: see also Fig.~4.10
of~\cite{Zinn97}.  However, in addition to the displacement $h_0(T,g)$, inclusion of
the surface tension background $\Sigma_0(t,h)$ will distort naive expectations
based on Fig.~\ref{fig4:Spm} when real isotherms for the total interfacial tension
are examined vs density or chemical potential.  See Fig.~\ref{fig4:isochore}(b)
below for another aspect of this issue.

It is appropriate to mention here that in their theoretical
analysis of a binary fluid mixture, NWW
\cite{NWW} tacitly \textit{assumed} symmetry in the surface tension above and below $T_c$
by supposing $S^+_M(\mt) = S^-_M(\mt)$: see their Eqs.~(2.8) and~(2.9) and Figs.~3 and ~10.
However, on physical grounds such a symmetry is quite implausible.
Thus, below $T_c$ there are two distinct fluid phases $\beta$
and $\gamma$ favored by $h>0$ and $h<0$, respectively, and a vapor
phase $\alpha$ (or wall) favoring the $\beta$ phase; but, above $T_c$,
there is only one fluid phase $\beta\gamma$: see Fig.~\ref{fig4:phase}.
Hence, one must allow for the $T \gl T_c$ symmetry breaking differences seen in Fig.~\ref{fig4:Spm}.
For comparison, one may recall that for the bulk equation of state when expressed in
terms of $\mt$ (or in terms of $\htilde$) one also needs \textit{two} scaling
functions, say $\thinspace\thinspace Q_\pm(\mt)$ for $t \ge 0$ and $t \le 0$, as is well
known; and, again, the two branches must be analytically related.  However, in the bulk
thermodynamics one may accept full asymptotic symmetry
under $\tilde h \Leftrightarrow -\tilde h$ or $\mt \Leftrightarrow -\mt$
and then employ just a single
function, say, in terms of the variable $t/|m|^{1/\beta} \propto \mt^{-1/\beta}$.
But that symmetry is also \textit{not} applicable to the surface tension.

Experimentally, binary mixtures are prepared at various fixed compositions and
the surface tension may then be observed as a function of temperature:
see, e.g.,~\cite{Pegg85}.  This is comparable
to keeping the order parameter $M$ fixed.  Accordingly we present such plots
for the EdGF predictions in Fig.~\ref{fig4:isochore}.

It must be recalled, however, that EdGF theory yields only the scaling or
singular part of the surface tension.  Since the leading power-law of the
surface tension, $|t|^\mu$, vanishes at the critical point, the contribution
from the analytic background is \textit{non-negligible}.  To illustrate this
point, we include in Fig.~\ref{fig4:isochore}(b) plots for the full surface tension,
$\Sigma = \Delta\Sigma + \Sigma_0$,
with an \textit{assumed} but reasonably realistic \textit{model} background term,
namely,
\begin{equation}
\Sigma_0(T,h)/K = 1 - 2 t + 5 t^2, \label{eq4.5:S0}
\end{equation}
where, for simplicity, only the
temperature dependence has been considered.  Since $K^+ < 0$ [see~(\ref{eq4.5:Ks})],
the scaling part of the surface tension above $T_c$ in zero-field, namely,
$\Delta\Sabr \approx K^+ |t|^\mu$, curves downwards as seen
in Fig.~\ref{fig4:isochore}(a).  However, in
Fig.~\ref{fig4:isochore}(b), the corresponding curvature in the full
surface tension now appears to be \textit{upwards} owing to the effects
of the background $\Sigma_0(T)$.
In fact, the same sign of apparent curvature is observed in the NWW experiment on isobutyric
acid and water \cite{NWW}; this clearly demonstrates the importance
of the background $\Sigma_0(T,h)$.
The significance of also introducing integral powers of $\tilde h$ in (\ref{eq4.5:SMpmnorm}) will be
discussed elsewhere~\cite{FisherNote}.

\subsection{The Complete-Wetting Singularity}
\begin{figure}
\includegraphics[scale=0.8]{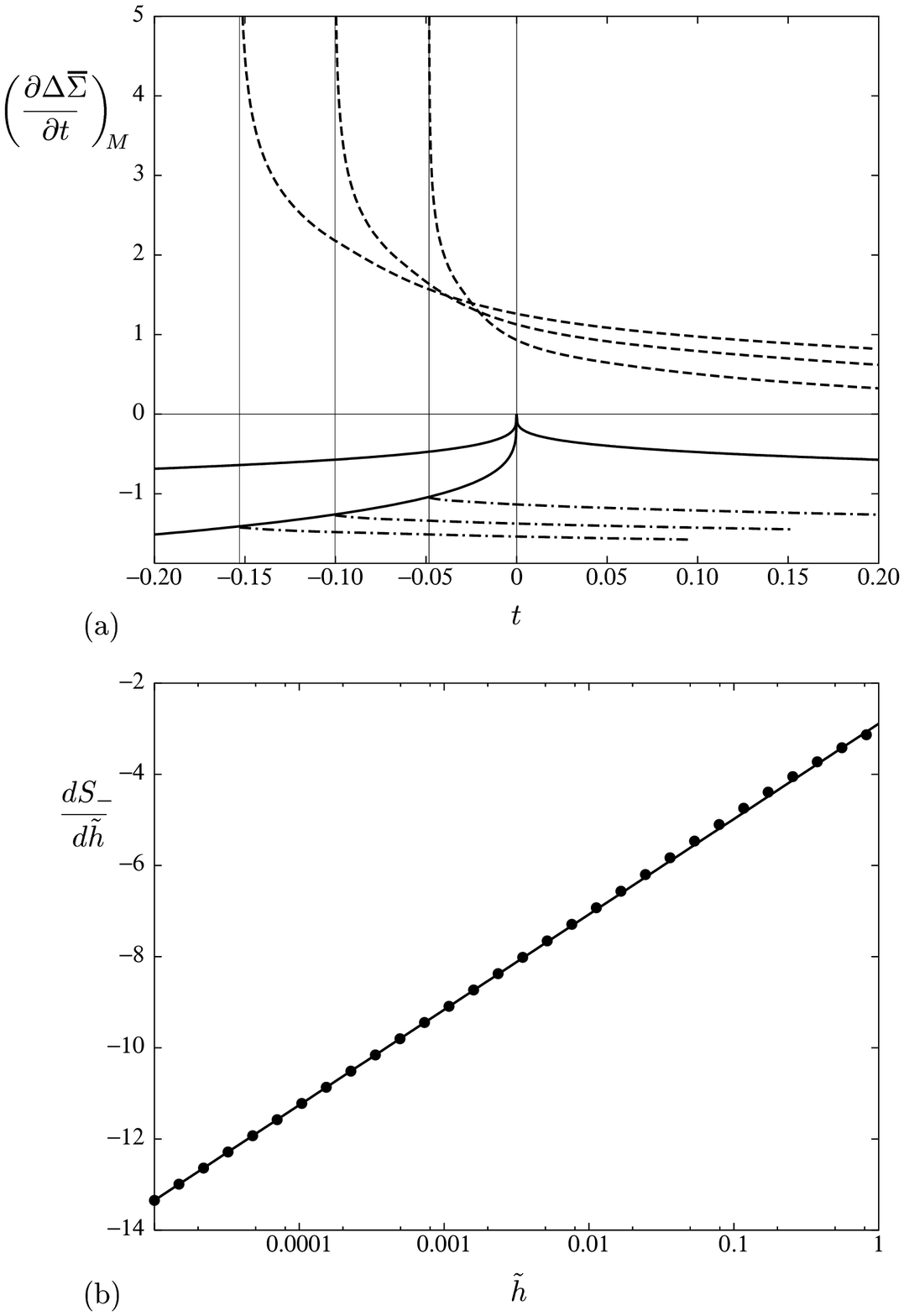}
\caption{\label{fig4:log}
(a) Temperature derivative of $\Delta{\overline \Sigma} \equiv
\Delta\Sigma/K$ at fixed $M$ (with values the same as in Fig.~\ref{fig4:isochore})
revealing a logarithmic singularity when $M$
is positive and $t\thinspace(<0)$ approaches the
coexistence curve; (b) semilogarithmic plot of the 
derivative of the scaling function $S_-(\tilde h)$ beneath
$T_c$: the straight line is a guide to the eyes
while the dots represent the derivative.}
\end{figure}

It has been predicted by Cahn~\cite{Cahn77} that a logarithmic singularity should
occur in the slope of $\Delta\Sar(T,h)$ on approaching the coexistence curve
below $T_c$, i.e., by taking
$\tilde h \to 0+$ (along any generic, nontangential path).
This logarithmic singularity is associated with the
complete-wetting transition that, in turn, is reflected in
Antonow's rule.  Recall that when $\tilde h=0$, the two liquid phases $\beta$ and
$\gamma$ coexist with the vapor phase $\alpha$.  Thus, when $\tilde h \to 0+$
while $\alpha$ and $\gamma$ phases coexist [i.e., on the surface $\sigma$
in Fig.~\ref{fig4:phase}(a)],
the $\beta$ phase of intermediate density emerges and spreads over (or wets)
the $\alpha | \gamma$ interface.
Indeed, one can explicitly show that such a singularity arises within the
EdGF theory: see Appendix C of~\cite{Zinn97}.
The analysis establishes that
the coefficient of the $\ln {\tilde h}_\infty$ term in $(\partial \Delta\Sigma/\partial \tilde h)$
is positive, which is fully consistent with the numerical calculations presented
in Fig.~\ref{fig4:log}: these demonstrate a $\ln (T-T_0)^{-1}$ singularity in the
derivative of the surface tension when $T$ approaches the coexistence curve at fixed
density $\rho > \rho_c$, i.e., $M > 0$.

Although the presence of a logarithmic singularity in $\Sar(h)$ on approaching a
wet interface may be regarded as well established theoretically,
an important caveat is that
in all cases the corresponding theoretical analysis entails the assumption that
interactions within the fluids are entirely of a short-range character, i.e., decaying
faster than any power-law.  This, of course, precludes slowly decaying interaction
potentials such as the $1/r^6$ form that characterizes the van der Waals intermolecular
forces prevalent in real molecular fluids.
At fixed temperature $T$ in the range $T_W < T < T_c$ van der Waals forces
should generate a $(h - h_0)^{-1/3}$ singularity in $\Sar(h)$.  Thus, without special
allowance for power-law potentials, a local functional analysis such as our EdGF theory
must be suspect in relation to real fluid systems unless short range interactions
happen to dominate for, say, reasons of symmetry, or accidental near-cancellation, etc.

On the other hand, in the case of $d=3$ bulk critical behavior it is known that $1/r^6$
potentials enter the asymptotic scaling forms only as \textit{irrelevant} corrections-to-scaling
(even though these may be \textit{dangerously} \textit{irrelevant} for certain quantities
such as correlation functions at long distances)~\cite{Kayser}.
It is possible that a similar situation pertains in the case of the surface tensions
near a critical endpoint in which case a local functional theory might still prove
asymptotically adequate.  To our knowledge, however, this issue remains open and,
as yet, a singularity in the derivative of the surface tension $\Sar(T,h)$ near a critical
endpoint has not been identified experimentally: see, e.g., \cite{NWW}.

\section{Summary}
As illustrated in Fig.~\ref{fig4:phase}, binary
fluid mixtures exhibit critical endpoints where, in the
three-dimensional
thermodynamic field space, a critical line of mixing transitions
terminates at a first-order transition surface between the liquid phases and
their common vapor phase.  At such critical
endpoints, the interfacial or surface tension $\Sigma(T,h)$ becomes singular
in both temperature and ordering field $h$.
Our aim here has been to calculate the scaling functions describing the
asymptotic behavior of the surface tensions through the whole neighborhood
of the critical endpoint.

To this end, the local functional theory of Fisher and Upton \cite{Fisher90a, Fisher90b}
in the extended de~Gennes-Fisher (EdGF) version of the theory, has been exploited
because it captures
many significant physical features
tied to the nonclassical values of the relevant critical exponents, especially $\eta > 0$.
As seen in~(\ref{eq4.4:Sigma}), the EdGF
theory requires suitable scaling representations for the auxiliary free
energy $W(M;T,\hinf,g)$ and for the correlation length factor
$\xi^2/2\chi$.  To generate these,
we have used the extended sine model of Ref.~9, 
since it embodies the appropriate
analytic behavior, extends smoothly through the two-phase region below $T_c$,
and fits the values of many important universal amplitude ratios.

However, both numerical and analytical~\cite{Zinn97} calculations lead to the
prediction of a small but unphysical cusp in the variation of the surface
tension $\Sigma(T,h)$ on crossing the critical isotherm at positive~$h$,
i.e., on entering the $\gamma$ region of the phase diagram: see Fig.~\ref{fig4:phase}(a).
This unanticipated behavior represents a shortcoming of the EdGF theory that
is found to originate in the predicted variation of the order parameter profile, $M(z)$,
in the immediate vicinity of $T=T_c$ when it passes through the critical value $M(z_0)=0$.
An improved local functional theory might avoid this difficulty.
To that end, it may be worthwhile to investigate the generalized or GdGF
theory~\cite{Fisher90b}.  However, the undesirable feature may still remain
since we are inclined to believe that the origin of the cusp is associated
rather directly with the single, scalar order-parameter formulation: both EdGF
and GdGF theories employ a simple scalar order parameter which cannot avoid ``local
criticality'' when the interfacial profile crosses from one phase to another at
$T \negthinspace = \negthinspace T_c$.  In response to this observation,
Mikheev and Fisher~\cite{Mik92} have addressed the formulation of two-order-parameter
theories in which, in particular, the local energy fluctuation, as a second `critical
density,' plays a role; but a practicable scheme of approximation has not so far
been achieved.

Nevertheless, the application of EdGF theory to other properties of fluid interfaces
and surfaces, seems worthwhile (e.g.~\cite{Upton2001}) and in the present case
the cusp in $\thinspace\thinspace s(\theta)$, the scaling function for the surface
tension, produces a deviation of only a few percent from the naturally interpolated
analytic variation.  Accordingly, for numerical purposes we have adopted a smoothing
procedure that removes the cusp; this yields the ameliorated angular function
${\overline s}(\t)$ that is recorded numerically in Table~I.

On this basis the universal scaling functions $S_\pm(\tilde h)$ and
$S^\pm_M(\mt)$ have been calculated: see Figs.~\ref{fig4:Spm} and~\ref{fig4:isochore} which
reveal significant features of the anticipated variation
of the surface tension $\Sigma(T,h)$.  In particular,
the role of the analytic background contribution
$\Sigma_0(T,h)$ can be assessed.  Together with
the improved theoretical predictions~(\ref{eq4.5:edgfPQ}) for the universal
surface tension amplitude ratios, $P$ and $Q$, these
results will be used elsewhere~\cite{FisherNote} to reanalyze the experimental
data of Nagarajan, Webb, and Widom~\cite{NWW} and to compare with other
experiments~\cite{Pegg85, Amara91, Mainzer}.

\begin{acknowledgments}
The interest of B.~Widom and P.J.~Upton has been appreciated
and we are indebted to H.W.~Diehl, S.~Dietrich, J.~Indekeu, and B.~Widom
for helpful comments on a draft manuscript.  The authors are
grateful for the support of the National Science Foundation through
grants CHE~99-81772 and CHE~03-01101.
\end{acknowledgments}

\bibliography{ms}%

\begin{thebibliography}{48}
\bibitem[{dim()}]{dimen}
\bibinfo{note}{It should also be noted that $M$, $h$, $A_s(M,T)$,
  $\xi^2/2\chi$, etc., now carry dimensions (in contrast to the formulation in
  Ref.~9) so that the corresponding angular functions $m(\t)$, $l(\t)$,
  $n(\t)$, $a_\infty(\t)$, etc., have appropriate dimensions. Consequently,
  $l_0$, $m_0$, and $a_{\infty,0} \equiv a_\infty(0)$, etc., also gain
  dimensions. However, the angular function $k(\t)$ for the reduced temperature
  is kept dimensionless as are the parametric variables $r$ and $\t$.}
\end{thebibliography}
\centerline{\vbox{\hrule width 3 in height 2 pt}}
\end{document}